\documentclass[sigconf]{acmart}
\usepackage{amsmath}
\usepackage{bm}
\usepackage{graphicx}
\usepackage{algorithm}
\usepackage{enumitem}
\usepackage[noend]{algorithmic}
\usepackage{multirow}
\usepackage{amsthm}
\usepackage{amsmath,stmaryrd,pict2e,picture}
\usepackage{amsmath}
\usepackage{appendix}

\DeclareMathOperator*{\argmin}{arg\,min}
\usepackage{subcaption}
\usepackage{hyperref}
\hypersetup{colorlinks,allcolors=black}

\newcommand{\ourmethod}{DiffPrep}
\newcommand{\stitle}[1]{\vspace{1ex}\noindent{\bf #1}}
\AtBeginDocument{%
  \providecommand\BibTeX{{%
    \normalfont B\kern-0.5em{\scshape i\kern-0.25em b}\kern-0.8em\TeX}}}

\setcopyright{acmlicensed}
\acmConference[SIGMOD '23]{2023 International Conference on Management of Data}{June 18--23,
  2023}{Seattle, WA}
\acmYear{2023} \acmVolume{1} \acmNumber{2} \acmArticle{183} \acmMonth{6} \acmPrice{}\acmDOI{10.1145/3589328}

\begin{document}

\title{\ourmethod: Differentiable Data Preprocessing Pipeline Search for Learning over Tabular Data}

\author{Peng Li}
\email{pengli@gatech.edu}
\affiliation{%
  \institution{Georgia Institute of Technology}
  \city{Atlanta}
  \state{GA}
  \country{USA}
}

\author{Zhiyi Chen}
\affiliation{%
  \institution{Georgia Institute of Technology}
  \city{Atlanta}
  \state{GA}
  \country{USA}
}
\email{zchen798@gatech.edu}

\author{Xu Chu}
\affiliation{%
  \institution{Georgia Institute of Technology}
  \city{Atlanta}
  \state{GA}
  \country{USA}
}
\email{xu.chu@cc.gatech.edu}

\author{Kexin Rong}
\affiliation{%
  \institution{Georgia Institute of Technology}
  \city{Atlanta}
  \state{GA}
  \country{USA}
}
\email{krong@gatech.edu}

\newcommand{\rev}[1]{{\color{black}#1}}
\newcommand{\todo}[1]{{\color{red}#1}}

\begin{abstract}
Data preprocessing is a crucial step in the machine learning process that transforms raw data into a more usable format for downstream ML models. However, it can be costly and time-consuming, often requiring the expertise of domain experts. Existing automated machine learning (AutoML) frameworks claim to automate data preprocessing. However, they often use a restricted search space of data preprocessing pipelines which limits the potential performance gains, and they are often too slow as they require training the ML model multiple times. In this paper, we propose \textsc{\ourmethod}, a method that can automatically and efficiently search for a data preprocessing pipeline for a given tabular dataset and a \rev{differentiable} ML model such that the performance of the ML model is maximized. We formalize the problem of data preprocessing pipeline search as a bi-level optimization problem. To solve this problem efficiently, we transform and relax the discrete, non-differential search space into a continuous and differentiable one, which allows us to perform the pipeline search using gradient descent with training the ML model only once.  Our experiments show that \textsc{\ourmethod} achieves the best test accuracy on \rev{15 out of the 18} real-world datasets evaluated and improves the model’s test accuracy by up to 6.6 percentage points.
\end{abstract}




\received{October 2022}
\received[revised]{January 2023}
\received[accepted]{February 2023}

\maketitle

\section{Introduction}
\label{sec:introduction}
\begin{figure}[t]
    \centering
    \includegraphics[width=0.8\columnwidth]{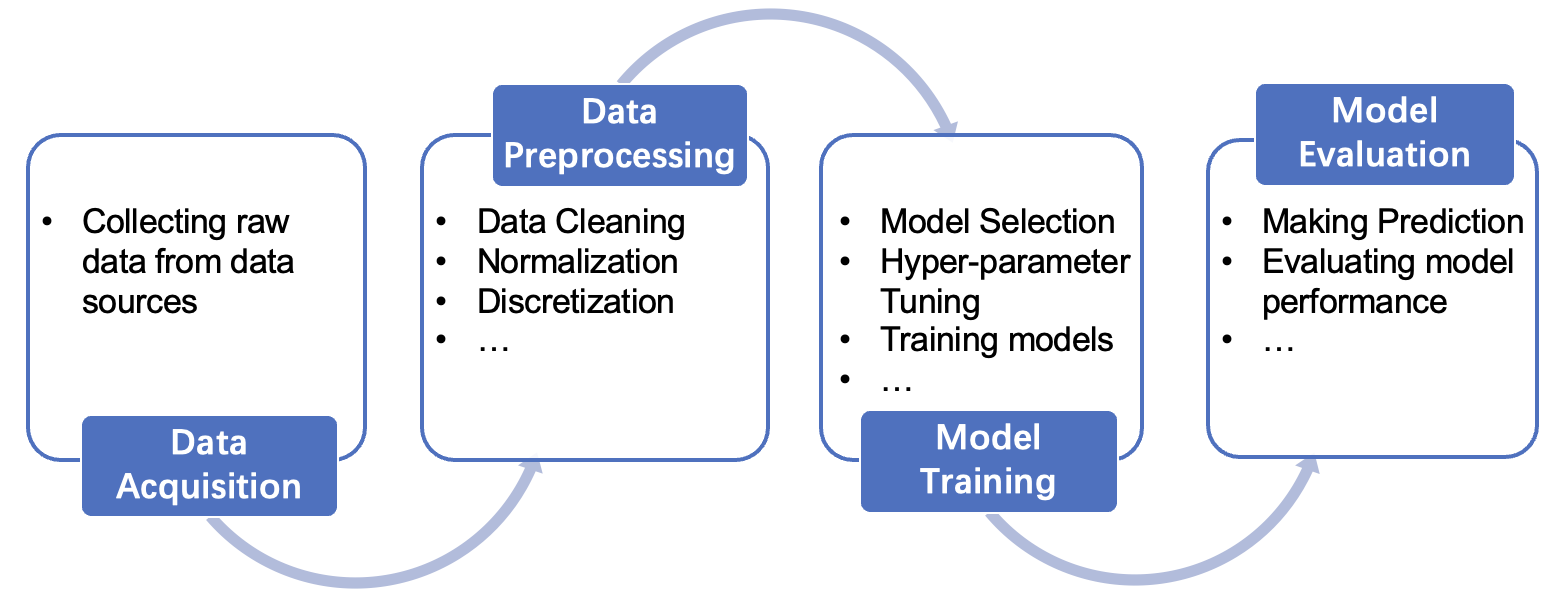}
    \caption{A Typical ML Development Workflow}
    \label{fig:ml_workflow}
\end{figure}

 Machine learning (ML), in particular, supervised ML is increasingly being used for solving challenging real-world problems in a wide range of fields, such as medicine~\cite{kourou2015machine}, finance~\cite{deng2016deep}, politics~\cite{ramteke2016election}, etc. The workflow of developing an ML application may vary from projects but it typically involves four stages as shown in Figure \ref{fig:ml_workflow}, including data acquisition, data preprocessing, model training, and model evaluation~\cite{he2021automl}.

Data preprocessing is an essential step in a typical ML workflow because in practice, the raw data collected often contain data issues and can rarely be directly used by ML models~\cite{garcia2015data}. For example, data errors such as missing values and outliers will significantly reduce the ML model performance if not cleaned~\cite{li2021cleanml}; certain models like  k-nearest neighbors (KNN)~\cite{peterson2009k} expect data in different columns to have a similar range, which requires the raw data to be normalized. In addition, real-world data often have multiple such data issues~\cite{abedjan2016detecting}: a dataset may contain missing values, outliers, and a large discrepancy on feature scales. To handle this case, we may first impute missing values with mean imputation, then remove outliers using the Z-score method, and finally normalize the data using standardization. As a result, data preprocessing usually involves multiple operators organized using a \textit{data preprocessing pipeline}, where the operators in the pipeline are applied sequentially and each operator transforms the data to tackle a specific data issue.

Designing data preprocessing pipelines is challenging for data scientists as it involves many design decisions on \textit{transformation types}, \textit{order}, and \textit{operators}~\cite{giovanelli2021effective, chu2016data}. First, data scientists must decide which types of transformations (e.g., outlier removal, discretization, normalization) are needed and the order of different transformations in the pipeline. For example, outlier removal may or may not be needed and can be applied before or after normalization. For each type of transformation, there are multiple choices of operators--for example, standardization, min-max scaling, and robust scaling are all commonly used operators for normalization. Data scientists need to decide which operator to use for each transformation. Furthermore, different features may need to be preprocessed differently, which requires data scientists to design a \textit{feature-wise pipeline} rather than using the same preprocessing pipeline for all features. For example, when repairing missing values in a dataset, some features may prefer mean imputation while others may prefer median imputation, depending on their distributions. 

Even for experienced data scientists, it is usually not clear how to design a preprocessing pipeline that will lead to the best performance. Making such decisions heavily relies on domain knowledge, including the characteristics of the data, the types of downstream ML models, and the data scientists' experience.
Traditional data cleaning works usually seek to design pipelines that optimize data quality independently of downstream applications~\cite{chu2016data}. However, since the ground-truth clean data of real-world datasets are rarely available, the data quality may not be accessible or accurately estimated. Moreover, previous works have shown that data cleaning or preprocessing  without considering downstream ML models can sometimes negatively impact the performance of ML models~\cite{li2021cleanml,neutatz2021cleaning}. In practice, data scientists often use the time-consuming trial-and-error method to design preprocessing pipelines, which is reported to account for 80\% of data scientists’ time~\cite{saha2014data}, or simply use some default configurations which usually result in suboptimal performance.

To reduce human effort in ML development, extensive study has been made on automated machine learning (AutoML) systems. Existing AutoML systems like Azure AutoML~\cite{Azure} and H2O.ai~\cite{ledell2020h2o} can automatically perform data preprocessing and model training without too much human involvement. Most AutoML systems consist of a search space and some optimization methods \cite{he2021automl}. The search space is defined by a set of possible parameters (choices) in ML development workflows. Our discussions in this paper focus on data preprocessing-related parameters, such as the choices of transformation types and operators. The optimization methods are used to automatically select a combination of parameters from the search space that leads to the highest model performance. With respect to data preprocessing pipelines, we find that existing AutoML solutions show limitations in both the size of search space and the efficiency of the optimization method:
\begin{figure*}[t]
\centering
\includegraphics[clip, trim=0.5cm 7cm 1cm 7cm, width=1.7\columnwidth]{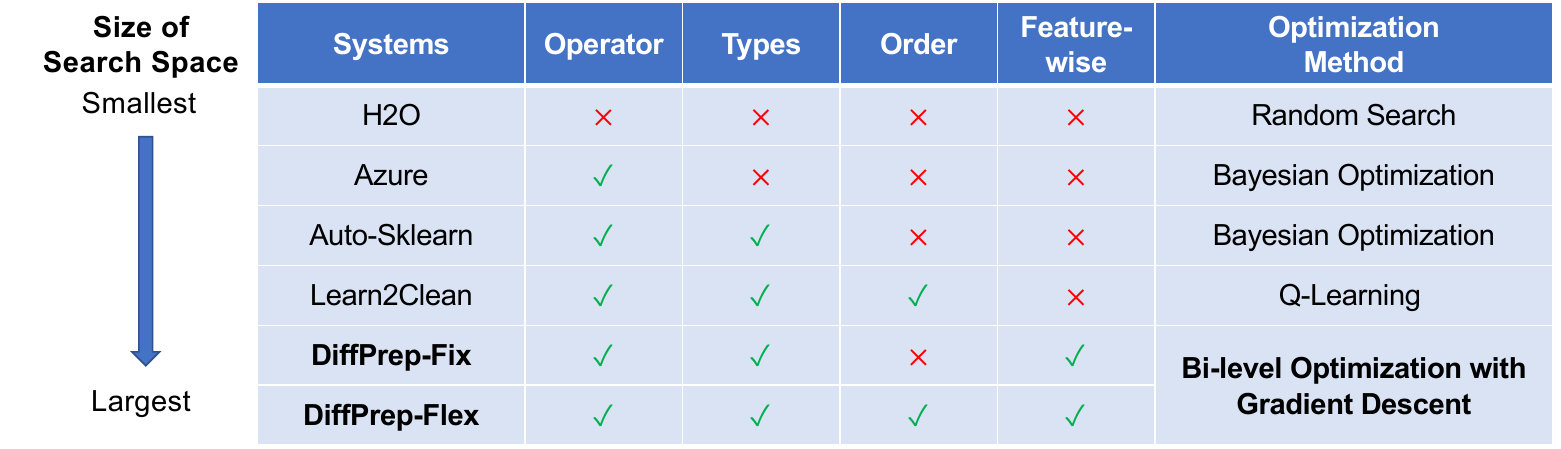}
\vspace{-2mm}
\caption{Data Preprocessing in AutoML Systems}
\label{fig:automl_data_preprocess}
\end{figure*}

\stitle{Limitation 1: Limited search space for data preprocessing pipelines.} The search space of an AutoML system determines the upper bound of its model performance. In general, larger search spaces are more likely to contain high-performing configurations, which could lead to better performance~\cite{ci2021evolving}. Despite the importance of data preprocessing in ML workflows, we found that existing AutoML frameworks  only consider a limited search space of parameters for data preprocessing pipelines. In Figure \ref{fig:automl_data_preprocess}, we rank existing AutoML systems according to the size of the search space they considered, which is determined by four variables: the choices of transformation types, the choices of transformation order, the choices of transformation operators, and whether the same pipeline is used for each data feature. H2O only supports a single default preprocessing pipeline with fixed transformation operators, types and order. Azure would search for the best operator for some transformations, such as normalization, but the transformation types and order are fixed.  Auto-Sklearn~\cite{feurer-neurips15a} can determine the types of transformation needed and select suitable operators, but it has a fixed order of applying transformations. Learn2Clean~\cite{berti2019learn2clean} considers operators, types and order in its search space, but it uses the same pipeline to preprocess all features. Therefore, existing AutoML systems have only explored a small portion of the entire design space of data preprocessing pipelines, which limits the performance gains of downstream ML models.

\stitle{Limitation 2: Low efficiency on optimization methods.} As the search space becomes larger, it is increasingly important that the optimization methods can efficiently identify parameters with good performance. Unfortunately, most existing AutoML systems use optimization methods like random search or Bayesian optimization (Figure \ref{fig:automl_data_preprocess}), which require training the ML model multiple times and do not scale well with the search space. Specifically, random search randomly samples parameters from the entire search space and trains the ML model with each set of sampled parameters; Bayesian optimization builds a probabilistic model that maps parameters to model performance, which requires iteratively training the ML model with new optimal parameters and updating the probabilistic model with the model performance.  The optimization becomes particularly challenging if we want to support a larger search space in data preprocessing, such as using feature-wise pipelines. For example, if there are $n$ possible pipelines for one feature, the number of possible pipelines for data with $c$ columns is $n^c$, which means the space grows exponentially with the number of features. When there are many features or the ML model is large, random search or Bayesian optimization can be computationally expensive and time-consuming. 

\stitle{Our proposal.} In this work, we propose \textsc{DiffPrep}, an automatic data preprocessing method that can efficiently select high-quality data preprocessing pipelines for any given dataset and differentiable ML model. 
Unlike traditional data preprocessing or cleaning methods that focus on improving data quality independently of the downstream applications~\cite{chu2016data}, \textsc{DiffPrep} co-optimizes data preprocessing with model training: the goal is \textit{to select data preprocessing pipelines that maximize the validation accuracy (or minimize the validation loss) of the ML model trained on the dataset.}  Since the model training aims at minimizing the training loss, while the pipeline selection aims at minimizing the validation loss,  we end up with a so-called \textit{bi-level optimization problem}~\cite{liu2021investigating}.

Different from existing AutoML systems, \textsc{DiffPrep} considers feature-wise data preprocessing pipelines in addition to transformation type, operators, and order. Feature-wise pipelines significantly broaden the search space for data preprocessing compared to existing AutoML systems, since the number of possible pipelines grows exponentially with the number of features. We consider two usage scenarios with \textsc{DiffPrep}. If users provide a pre-defined order, our method (\ourmethod-Fix) will fix the order and automatically select the types and operators to generate a pipeline for each feature. If no order is provided, our method (\ourmethod-Flex) will automatically select the  order, type and operator to generate a pipeline for each feature. It would explore the entire design space of data preprocessing pipelines, which is the largest search space in Figure \ref{fig:automl_data_preprocess}. In both scenarios, we have empirically found that the larger search space improves the quality of the resulting pipelines.

Efficiently searching over this large search space is challenging: since the search space is discrete and non-differentiable, we need to enumerate each option and train ML models repeatedly to find high-quality pipelines. Our key insight is to make the ML model performance differentiable with respect to preprocessing pipeline choices so that we can leverage efficient optimization methods like gradient descent. To do so, we first parameterize the search space of data preprocessing pipelines such that each choice in the pipeline can be represented using binary parameters. We then relax the search space to be continuous using the softmax function and Sinkhorn normalization to make the pipeline differentiable. This allows \textsc{DiffPrep} to use gradient descent as the optimization method to solve the bi-level optimization problem, which allows us to optimize the pipeline and model simultaneously with training the ML model only once. 

 \stitle{Contributions.} We make the following contributions in this paper.
 \begin{itemize}[leftmargin=*,topsep=0pt]
  \item  We propose \textsc{DiffPrep}, the first automatic data preprocessing method to consider the design space of transformation types, operators, order and feature-wise pipelines. 
  \item We formalize the problem of automatic data preprocessing as a bi-level optimization problem and use gradient descent to solve the bi-level optimization problem efficiently.
    \item We conduct experiments on \rev{18} real-world datasets to evaluate the effectiveness of our method. The results show that our method achieves the best test accuracy on \rev{15 out of 18} datasets and improves the test accuracy by up to 6.6 percentage points.
  \end{itemize}

 \stitle{Organization.}
The rest of this paper is organized as follows. Section \ref{sec:preliminary} introduces preliminaries for constructing a data preprocessing pipeline and formally defines our studied problem. In Section \ref{sec:auto_prep_fixed_prototype}, we discuss our method to automatically search a data preprocessing pipeline given a fixed order of transformations. In Section \ref{sec:autoprep_flex_prototype}, we present our approach with flexible order of transformations. In Section \ref{sec:experiment}, we show the experimental results. We discuss the related work in Section \ref{sec:related_work} and conclude the paper in Section \ref{sec:conclusion}.

\section{Preliminary}
\label{sec:preliminary}

\subsection{Transformation Operators and Types}
\label{sec:tf_operators}
\stitle{Transformation Operators.} Generally, a data preprocessing operator/algorithm, which we term as a \textit{transformation operator (TF operator)}, is a function defined by $f : \mathcal{X}_1 \rightarrow \mathcal{X}_2 $,
where an original feature in feature space $\mathcal{X}_1$, is mapped to a transformed feature in feature space $\mathcal{X}_2$. In this paper, we focus on TF operators where both input and output are scalars, which we refer to as \textit{feature-wise transformation operators}. Other TF operators that transform a vector into another vector, such as principal component analysis, and entity resolution operators~\cite{cappuzzo2020creating,zhao2022leva} that take a vector as input, are not considered in this paper. We note that most TF operators used in practice fall into this category. For example, among the TF operators provided by \textit{scikit-learn}~\cite{scikit-learn} in \texttt{sklearn.preprocessing} module, 14 out of 18 are feature-wise TF operators. Some TF operators have parameters. For example, to use the Z-Score method, we need to specify the threshold $k$ to determine whether a value is an outlier (e.g., k = 3). 

\stitle{Transformation Types.} Based on the purpose of the transformation, TF operators can be grouped into different types, which we refer to as \textit{transformation types (TF types)}. For example, missing value imputation is a TF type that consists of TF operators such as mean imputation and median imputation. Table \ref{tab:preprocessing_space} shows some examples of TF types and the TF operators for each TF type. Formally, we define a TF type $F$ as a set of TF operators, denoted by \rev{$F = \{f_i : i= 1, 2, ...\}$}. 

\rev{\stitle{Design Considerations for Transformation Operators.} The possibilities for transformation operators are much greater than what we cover in this paper. For example, it is possible to define custom conditions for selecting specific subsets of the dataset and to use custom functions to transform them. However, it is difficult to search in such a vast search space. Furthermore, when co-optimizing data preprocessing and model training, using a large number of operators that can modify data \emph{arbitrarily} can increase the risk of model overfitting, as observed in prior work~\cite{krishnan2017boostclean}. Therefore, \textsc{DiffPrep} focuses on a limited set of operators that are commonly supported by ML frameworks (e.g. scikit-learn) and widely used. We have found empirically that this set of operators already leads to improved performance on many real-world datasets. Additionally, the operators supported by \textsc{DiffPrep} are designed for general purpose to transform data based on some prior knowledge rather than modifying the dataset arbitrarily. For instance, operators for outlier removal and missing value imputation can only affect specific (usually small) subsets of the dataset, which are usually prone to be dirty data, while normalization and discretization operators can only scale and shift the distribution of entire features, but not completely distort the data distribution. }

\begin{table}[t]
\centering
\caption{Example of TF Types and TF operators}
\vspace{-2mm}
\scalebox{0.8}{
\begin{tabular}{|c|cc|}
\hline
\textbf{Transformation Types} & \multicolumn{2}{c|}{\textbf{Transformation Operators}} \\ \hline
\multirow{5}{*}{\begin{tabular}[c]{@{}c@{}}Missing Value \\ Imputation\end{tabular}} & \multicolumn{1}{c|}{\multirow{3}{*}{\begin{tabular}[c]{@{}c@{}}Numerical \\ Features\end{tabular}}} & Mean \\
 & \multicolumn{1}{c|}{} & Median \\
 & \multicolumn{1}{c|}{} & Mode \\ \cline{2-3} 
 & \multicolumn{1}{c|}{\multirow{2}{*}{\begin{tabular}[c]{@{}c@{}}Categorical \\ Features\end{tabular}}} & Most Frequent Value \\
 & \multicolumn{1}{c|}{} & Dummy Variable \\ \hline
\multirow{4}{*}{Normalization} & \multicolumn{2}{c|}{Standardization} \\
 & \multicolumn{2}{c|}{Min-Max Scaling} \\
 & \multicolumn{2}{c|}{Robust Scaling} \\
 & \multicolumn{2}{c|}{Max Absolute Scaling} \\ \hline
\multirow{3}{*}{\begin{tabular}[c]{@{}c@{}}Outlier \\ Removal\end{tabular}} & \multicolumn{2}{c|}{Z-Score $(k)$} \\
 & \multicolumn{2}{c|}{MAD $(k)$} \\
 & \multicolumn{2}{c|}{IQR $(k)$} \\ \hline
\multirow{2}{*}{Discretization} & \multicolumn{2}{c|}{Uniform $(n)$} \\
 & \multicolumn{2}{c|}{Quantile $(n)$} \\ \hline
\end{tabular}
}    
\label{tab:preprocessing_space}
\end{table}

\subsection{Data Preprocessing Pipeline Construction}
\label{sec:data_pipeline_construct}

\begin{figure*}[t]
    \centering
    \includegraphics[width=1.8\columnwidth]{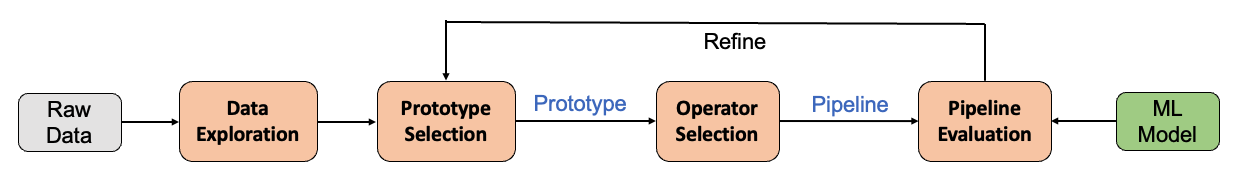}
    \vspace{-4mm}
    \caption{A typical data scientists' workflow to construct a data preprocessing pipeline}
    \label{fig:data_preprocessing_process}
\end{figure*}

As we mentioned before, data preprocessing usually involves multiple transformations that are combined in a pipeline and each feature can use a different pipeline. For simplicity, let us first assume that the data only contain one feature and we want to construct a data preprocessing pipeline for it. Figure \ref{fig:data_preprocessing_process} shows a typical data scientists' workflow to construct a data preprocessing pipeline, which involves the following steps.

\stitle{Step 1: Data exploration.} The initial step in creating a data preprocessing pipeline is usually \textit{data exploration}. During this step, data scientists examine the data to understand its characteristics and identify any potential or existing issues. Some issues, such as missing values, are relatively easy to detect, while others, such as outliers, may require more involved data analysis~\cite{hawkins2002outlier}.

\stitle{Step 2: Prototype selection.} The second step is called \textit{prototype selection}, where based on the data issues, data scientists would select the TF types to be involved and decide the order of TF types to be applied~\cite{quemy2020two, giovanelli2021effective,shang2019democratizing}. For example, we choose to first impute missing values, then remove outliers and finally perform data normalization. Note that in a pipeline, the input of one operator is the output of its previous operator. Therefore, different orders of TF types can result in totally different outputs. For example, if outlier removal occurs before normalization, not only the input data for normalization are changed, but also the statistics used to normalize the data (e.g., the minimum and maximum value of the column) are affected due to the removal of outliers, which can yield a significantly different output compared with having outlier removal after normalization. However, in practice, it is usually not clear how to decide the order of TF types and  data scientists would simply choose some default order based on their experience. The outcome of this step is a \textit{data preprocessing prototype}, which is an ordered sequence of TF types formally defined as follows. 

\begin{definition} (Data Preprocessing Prototype).
Let $\mathcal{S} = \{F\}$ denote the space of TF types. We define a data preprocessing prototype $\mathcal{T}$ as $\mathcal{T} = \{T_i \in \mathcal{S} : i = 1, 2, ...\}$, where $T_i$ is the $i$-th TF type to be applied and selected from the space. 
\end{definition}

Note that in the literature, the notion of data preprocessing prototype is also known as logical pipeline plan~\cite{shang2019democratizing} or pipeline prototype~\cite{quemy2020two}, which is defined as a directed acyclic graph of TF types. This definition suggests that there is no repetition of a TF type in a prototype~\cite{giovanelli2021effective}. This is because in practice, a type of transformation would rarely be repeated (e.g., it does not make sense to impute missing values twice). Following this widely-used definition of prototype, we require that $ \forall p \neq q, T_p \neq T_q$. We want to point out that this assumption does not prevent users from using a TF type multiple times. If repetition is really needed, we can define two TF types $T_1$ and $T_2$ on the same transformation and consider them as two different TF types. This allows for repeated transformation types in a prototype. 

\stitle{Step 3: Operator selection.} The third step is \textit{operator selection}, where given a data preprocessing prototype, data scientists would select a specific TF operator for each TF type in the prototype. For example, we use mean imputation to impute missing values, use the Z-score approach to remove outliers and perform standardization to normalize data. To select suitable operators for TF types, data scientists have to consider not only the characteristics of data but also the downstream ML model. For example, for normalization, if the downstream model is k-nearest neighbors, min-max scaling is usually preferable as it transforms all the columns into the same scale. In contrast, if the model is logistic regression, standardization may be better as it makes convergence faster. However, such heuristic rules may not work for every dataset,  and in practice, selecting suitable operators usually requires trial-and-error. By selecting a TF operator for each TF type, we can instantiate a \textit{data preprocessing pipeline}, which is formally defined as follows.

\begin{definition} (Data Preprocessing Pipeline).
Given a data preprocessing prototype $\mathcal{T}$,  we define a data preprocessing pipeline $\mathcal{G}_{\mathcal{T}}$ as $\mathcal{G}_{\mathcal{T}} = \{g_i: i = 1, 2, ... \}$, where $g_i \in T_i$ is the TF operator selected for $T_i$ in the prototype and the $i$-th operator to be applied in the pipeline.
\end{definition}

\stitle{Step 4: Pipeline evaluation.} The final step is \textit{pipeline evaluation}, where data scientists would use the pipeline to transform the raw data and evaluate the performance of the pipeline by training and testing the end ML model on the transformed data. To transform the raw data with the pipeline, we can  sequentially apply the TF operator in the pipeline. Formally, let $x$ denote the raw feature of one example and $x_i$ be the transformed feature after $i$ steps transformation, where $x_0 = x$. Then, $x_i$ can be computed by applying $g_i$ (the $i$-th TF operator in the pipeline)  on its input $x_{i-1}$ as: $x_i = g_i(x_{i-1})$.

The output of the pipeline is the output of the last TF operator in the pipeline. Formally,
let $s$ denote the number of TF operators in the pipeline and $\mathcal{G}_{\mathcal{T}}(x)$ denote the   final output of the pipeline $\mathcal{G}_{\mathcal{T}}$. Then, we have $\mathcal{G}_{\mathcal{T}} (x) = x_s$.

After pipeline evaluation, data scientists may refine the prototype and the pipeline based on the evaluation results. For example, they may add/delete TF types in the prototype, change the order of TF types, replace the TF operator for some TF types, etc. The above process will be repeated until the end ML model achieves desired performance. As a result, constructing data preprocessing pipelines is an iterative process that requires substantial domain knowledge and heavily relies on human experts to make decisions, which can be time-consuming and costly.

\subsection{Problem Statement}
\label{sec:problem_statement}
The core idea of our approach is to formulate the decision-making process of data scientists as an optimization problem. 
Assume that each data example has $c$ features denoted by $\bm{x} = [x^1, x^2, ... x^c]$. Let $\mathcal{T}^i, \mathcal{G}_{\mathcal{T}}^i$ denote the prototype and the pipeline for the $i$-th feature. Let $\bm{w}$ denote the parameters of the ML model $h_{\bm{w}} : \mathcal{X}^c \rightarrow \mathcal{Y}$. The model will take features of an example as input and generate a prediction $\hat{y} = h_{\bm{w}}(x^1, x^2, ... x^c)$. Let $loss(\hat{y}, y)$ be the loss function that returns a loss score given the prediction $\hat{y}$ and the ground truth label $y$. Then, the training loss and validation loss on the transformed data can be computed as:
$$L_{train}(\mathcal{G}_{\mathcal{T}}^1,..., \mathcal{G}_{\mathcal{T}}^c, \bm{w}) = \sum_{\bm{x}, y \in D_{train}} loss(h_{\bm{w}}(\mathcal{G}_{\mathcal{T}}^1(x^1),..., \mathcal{G}_{\mathcal{T}}^c(x^c)), y) $$
\vspace{-1mm}
$$L_{val}(\mathcal{G}_{\mathcal{T}}^1,..., \mathcal{G}_{\mathcal{T}}^c, \bm{w}) = \sum_{\bm{x}, y \in D_{val}} loss(h_{\bm{w}}(\mathcal{G}_{\mathcal{T}}^1(x^1),..., \mathcal{G}_{\mathcal{T}}^c(x^c)), y)$$
\vspace{-2mm}

\stitle{Problem Statement.} Given a training set $D_{train}$ and a validation set $D_{val}$, a space of TF operators and TF types $\mathcal{S}$, and a set of ML model parameters $\bm{w}$, we would like to find a pipeline $\mathcal{G}_{\mathcal{T}}^i$ (with its prototype $\mathcal{T}^i$) from the space for each feature $x^i$, such that the performance of the ML model trained and evaluated on the transformed data is maximized. This \textit{data preprocessing pipeline search problem (DPPS)} can be formulated as an optimization problem:
\begin{align*}
\min_{\mathcal{G}_{\mathcal{T}}^1,..., \mathcal{G}_{\mathcal{T}}^c} \quad & L_{val}(\mathcal{G}_{\mathcal{T}}^1,..., \mathcal{G}_{\mathcal{T}}^c, \bm{w}^*)\\
\vspace{-1mm}
\text{s.t.} \quad & \bm{w}^*= \argmin_{\bm{w}} L_{train}(\mathcal{G}_{\mathcal{T}}^1,..., \mathcal{G}_{\mathcal{T}}^c, \bm{w})   
\end{align*}
This is called a \textit{bi-level optimization problem} in which one optimization problem is embedded within another~\cite{liu2021investigating}. \rev{In the inner optimization, we fix the preprocessing pipeline parameters $\mathcal{G}_{\mathcal{T}}^1,..., \mathcal{G}_{\mathcal{T}}^c$ and focus on finding the best model parameters $\bm{w}^*$ to minimize the training loss on the transformed training data. In the outer optimization, we fix the model parameters and focus on finding the best pipeline parameters to minimize the validation loss on the transformed validation data.} \rev{An alternative problem formulation is to optimize both the pipeline and model parameters to minimize the training loss, which becomes a one-level optimization. However, previous works (e.g., DARTS~\cite{liu2018darts}) have shown that one-level optimization would make it easier to overfit the training data. 
Therefore, in this work, we use bi-level optimization to reduce the risk of overfitting. In our experiments, we empirically verified that the bi-level approach helps improve the model test accuracy compared to using a one-level optimization (Section~\ref{eval:ablation}).}

The most straightforward way to solve the above bi-level optimization problem is to train the downstream ML model by minimizing $L_{train}(\mathcal{G}_{\mathcal{T}}^1,..., \mathcal{G}_{\mathcal{T}}^c, \bm{w})$ for each possible data preprocessing pipeline in the space, and then select the one that has the minimal validation loss $L_{val}(\mathcal{G}_{\mathcal{T}}^1,..., \mathcal{G}_{\mathcal{T}}^c, \bm{w}^*)$. This naive approach is computationally expensive as it requires training the downstream ML models as many times as the number of possible data preprocessing pipelines. Consider a prototype with $s$ TF types, where each TF type consists of $m$ TF operators. Under this prototype, there are $m^s$ possible pipelines by choosing different TF operators for each TF type. Since different features can use different pipelines, with $c$ features, we have $m^{sc}$ possible pipelines, which is exponential to the number of TF types and the number of features. This is only the number of possible pipelines under one prototype. If we consider using different prototypes, the space will become even larger. Therefore, this naive approach is infeasible in practice.

\section{Data Preprocessing with Fixed Prototype} \label{sec:auto_prep_fixed_prototype}

Let us first assume that we have a fixed pre-defined data processing prototype $\mathcal{T} = \{T_1, ...,T_s\}$ that is used for all features. This is a common scenario where data scientists would like to skip the prototype selection step and use a default prototype so that they can spend more time on operator selection. This is also the setup of many existing systems~\cite{quemy2020two,thornton2013auto, feurer2020auto, berti2019learn2clean} for automatic data preprocessing, where the prototype is pre-defined by users and fixed in advance. 

Given this prototype, we need to assign each TF type in the prototype with a TF operator to generate a pipeline and we need to generate a pipeline for each feature. As the number of possible choices is exponentially large, the problem is: \textit{how to efficiently find the optimal assignment for each feature such that the validation loss is minimized?} This search problem is challenging because the search space is discrete and non-differentiable, which means we have to enumerate every possible case to find the optimal one. To solve this problem, we first parameterize the search space such that each assignment of TF operators (i.e., each choice of pipelines) can be represented using binary pipeline parameters (Section~\ref{sec:fixed_param}). Then, we relax the search space to be continuous so that the model loss is differentiable w.r.t the pipeline parameters (Section~\ref{sec:fixed_relax}). The relaxation enables us to solve the bi-level optimization problem efficiently using gradient descent (Section~\ref{sec:fixed_optimization}). 

\begin{figure*}[t]
 \centering
 \includegraphics[width=1.9\columnwidth]{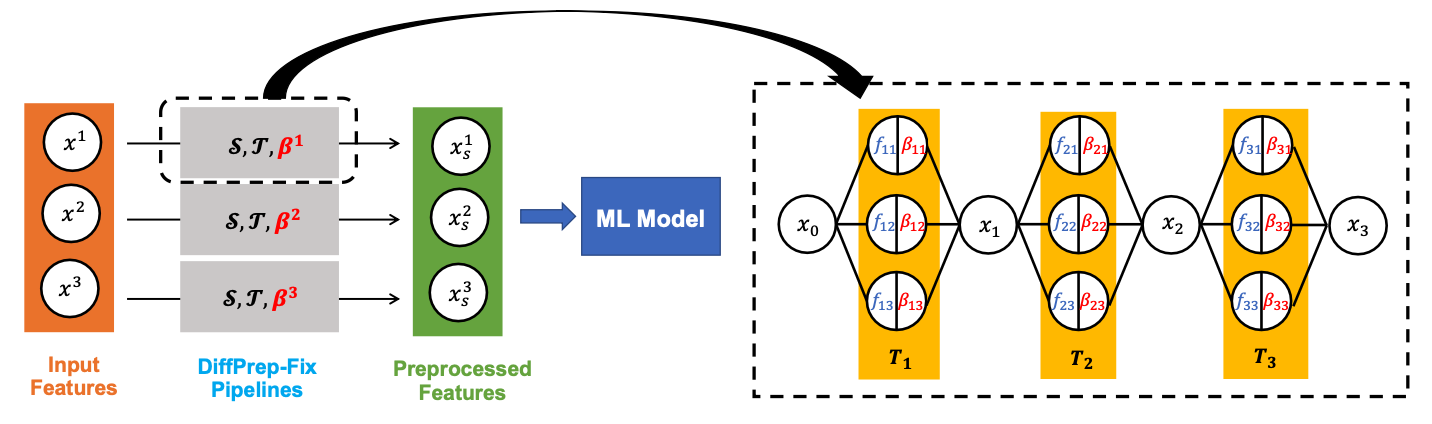}
 \vspace{-2mm}
 \caption{\ourmethod-Fix Architecture}
 \label{fig:autoprep_architect}
\end{figure*}

\subsection{Parameterization}
\label{sec:fixed_param}

Without loss of generality, assume that each TF type $T_i$ in the prototype consists of $m$ TF operators denoted by $T_i = \{f_{i1}, f_{i2}, ... f_{im}\}$ (if different TF types contain a different number of operators, $m$ denotes the maximum number of TF operators). To instantiate a pipeline, we need to select one specific TF operator for each TF type. Each selection can be represented using a $s \times m$ matrix $\bm{\beta} = \{\beta_{ij}\}, \beta_{ij} \in \{0, 1\}$ is defined as follows:
\begin{equation}
    \beta_{ij} =
    \begin{cases}
        1 & f_{ij} \text{ is selected}\\
        0 & \text{Otherwise}
    \end{cases}
    \label{eqn:def_beta}
\end{equation}
In other words, $\beta_{ij}$ is 1 if we select $f_{ij}$ for $T_i$, otherwise it is 0. Note that because only one TF operator is selected for each TF type, there is exactly one element each row in the matrix to be 1 and 0s elsewhere. Hence, we have $\sum_j \beta_{ij} = 1$. 

\begin{example}
Assume that the given prototype consists of three TF types, i.e., $\mathcal{T} = \{T_1, T_2, T_3\}$. Each TF type consists of four possible TF operators, i,e., $T_i = \{f_{i1}, f_{i2}, f_{i3}, f_{i4}\}$. Consider the data preprocessing pipeline $\mathcal{G}_{\mathcal{T}}=\{f_{12}, f_{21}, f_{34}\}$. The selection of TF operators that generates this pipeline can be represented using the following matrix.
\begin{equation*}
\bm{\beta} = \begin{bmatrix}
0 & 1 & 0 & 0\\ 
1 & 0 & 0 & 0\\ 
0 & 0 & 0 & 1\\ 
\end{bmatrix}
\end{equation*}
\end{example}

Each $\bm{\beta}$ matrix uniquely defines a data preprocessing pipeline under the given prototype and we can use $\bm{\beta}$ matrix to compute the output of the pipeline. Let $x$ denote the raw feature of one example and $x_i$ be the transformed feature after $i$ steps transformation, where $x_0 = x$. Then, we can compute $x_i$ as:
\begin{equation}
\label{eqn:forward_prop}
x_i = \sum^m_{j=1}\beta_{ij} f_{ij}(x_{i-1})
\end{equation}
Since only $\beta_{ij}$ associated with the selected TF operator is equal to 1 and others are 0, Equation~\ref{eqn:forward_prop} returns exactly the output of the selected TF operator. The final output of the pipeline $x_s$ becomes a variable parameterized by $\bm{\beta}$ matrix denoted by  $x_s=x(\bm{\beta})$. Assume that the dataset has $c$ features and let $\bm{\beta}^i$ denote the parameters that defines the pipeline for the $i$-th feature $x^i$. Then, the training loss and the  validation loss also become variables parameterized by all $\bm{\beta}$ matrices, which can be represented as:
\begin{equation}
    L_{train}(\bm{\beta}^1, ...,\bm{\beta}^c, \bm{w}) = \sum_{\bm{x}, y \in D_{train}} loss(h_{\bm{w}}(x^1(\bm {\beta^1}), ...,x^c(\bm {\beta^c})), y)
    \label{eqn:train_loss_param}
\end{equation}
\vspace{-2mm}
\begin{equation}
    L_{val}(\bm{\beta}^1, ...,\bm{\beta}^c, \bm{w}) = \sum_{\bm{x}, y \in D_{val}} loss(h_{\bm{w}}(x^1(\bm {\beta^1}), ...,x^c(\bm {\beta^c})), y)
    \label{eqn:val_loss_param}
\end{equation}
We can now rewrite the DPPS problem statement with a fixed prototype using $\bm{\beta}$ matrices as follows, where the search space of pipelines is converted into a space of $\bm{\beta}$ matrices. 
\begin{align}
\label{eqn:problem_statement_param_outer}
\min_{\bm{\beta}^1,..., \bm{\beta}^c} \quad & L_{val}(\bm {\beta}^1,... \bm {\beta}^c, \bm{w}^*)\\
\label{eqn:problem_statement_param_inner}
\text{s.t.} \quad & \bm{w}^* = \text{argmin}_{\bm{w}}  L_{train}(\bm{\beta}^1,..., \bm{\beta}^c, \bm{w}) 
\end{align}

 \stitle{Architecture.} Figure \ref{fig:autoprep_architect} shows the architecture of DiffPrep with a fixed prototype. For each input feature, DiffPrep will learn a set of $\bm{\beta}$ parameters that defines a data preprocessing pipeline. Each $\beta_{ij}$ parameter is attached with a TF operator indicating whether this operator is selected. The final output of DiffPrep pipelines will be the preprocessed features parameterized by $\bm{\beta}$ parameters, which will be fed into ML models for training and evaluation.
 
\subsection{Differentiable Relaxation}
\label{sec:fixed_relax}
Although we have parameterized the search space of pipelines using $\bm{\beta}$ matrices, the search space is still discrete as $\bm{\beta}$ matrices are binary parameters, i.e., $\beta_{ij} \in \{0, 1\}$. To make the search space continuous, we relax $\bm{\beta}$ to be continuous parameters that can take continuous values from 0 to 1, i.e., $\beta_{ij} \in [0, 1]$. However, we need to retain the constraints between parameters in the original binary matrix ($\sum_j \beta_{ij} = 1$) such that these parameters are semantically meaningful. To enforce that each row sums up to 1, we can define $\beta_{ij}$ using a softmax function as:
\begin{equation}
\label{eqn:beta_definition}
    \beta_{ij} =  \frac{\exp(\tau_{ij})}{\sum_k{\exp(\tau_{ik})}}
\end{equation}
where $\bm{\tau} = \{\tau_{ij}\} \in \mathbb{R}^{s \times m}$ are underlying parameters. 

By enforcing these constraints, we can interpret  $\beta_{ij}$ (Equation \ref{eqn:def_beta}) as the probability that $f_{ij}$ is selected for $F_i$. We can still use Equation \ref{eqn:forward_prop} to compute the transformed data, which can be interpreted as  the expected value of the transformed data. Similarly, we can compute the expected value of training loss and validation loss using Equation \ref{eqn:train_loss_param} and Equation \ref{eqn:val_loss_param}.

\subsection{Bi-level Optimization}
\label{sec:fixed_optimization}
Making $\bm {\beta}$ continuous allows us to solve the bi-level optimization problem  efficiently using gradient descent instead of enumerating all possible $\bm {\beta}$. To minimize the validation loss (Equation \ref{eqn:problem_statement_param_outer}), we can iteratively update the underlying parameters $\bm{\tau}$ using the gradient of the validation loss with respect to $\bm{\tau}$ as follows:
\begin{equation}
    \bm{\tau} = \bm{\tau}- \eta_1  \nabla_{\bm{\tau}} L_{val}( \bm{\beta}(\bm{\tau}), \bm{w}^*)
\label{eqn:gradient_descent_tau}
\end{equation}
where $\eta_1$ is the learning rate for updating $\bm{\tau}$. However, to obtain the optimal model $\bm{w^*}$ in Equation \ref{eqn:gradient_descent_tau}, we need to solve the inner optimization problem (Equation \ref{eqn:problem_statement_param_inner}) by completely training an ML model until convergence for every update of $\bm{\tau}$, which is computationally expensive. To solve this issue, instead of finding the optimal $\bm{w}^*$, we approximate it by doing only a single training step, which is a one-step gradient descent on the current model parameters $\bm{w}$ (denoted by $\bm{w}'$) as:
\begin{equation}
\bm{w}^* \approx \bm{w'} = \bm{w} - \eta_2\nabla_{\bm{w}}L_{train}( \bm{\beta}(\bm{\tau}), \bm{w})
\label{eqn:approx_optimal_model}
\end{equation}
where $\eta_2$ is the learning rate for model training~\footnote{In practice, to avoid tuning two learning rates, we can simply set $\eta_1 = \eta_2$~\cite{liu2018darts}.}. Similar approximation can also be found in previous work~\cite{franceschi2017forward, liu2018darts} for solving other bi-level optimization problems. We summarize the procedure of solving the bi-level optimization (Equation \ref{eqn:problem_statement_param_outer} and Equation \ref{eqn:problem_statement_param_inner}) using gradient descent in Algorithm \ref{alg:bilevel_gradient_descent}, where the pipeline parameters $\bm{\beta}(\bm{\tau})$  and model parameters $\bm{w}$ are alternately updated using gradient descent until convergence. 
\begin{small}

\begin{algorithm}[t]
\caption{Solving Bi-level Optimization with Gradient Descent}
\label{alg:bilevel_gradient_descent}
\begin{algorithmic}[1]
\STATE{Initialize $\bm{\tau}$ and $\bm{w}$}
\WHILE{\textit{not converged}}
\STATE{Update $\bm{\tau}$: $\bm{\tau} = \bm{\tau}- \eta_1  \nabla_{\bm{\tau}} L_{val}(\bm{\beta} (\bm{\tau}),  \bm{w} - \eta_2\nabla_{\bm{w}}L_{train}( \bm{\beta}(\bm{\tau}), \bm{w}))$}

\STATE{Update $\bm{w}$: $\bm{w} = \bm{w} - \eta_2\nabla_{\bm{w}}L_{train}( \bm{\beta} (\bm{\tau}), \bm{w})$}

\ENDWHILE

\RETURN $\bm{\beta}(\bm{\tau})$, $\bm{w}$
\end{algorithmic}
\end{algorithm}
    
\end{small}

\stitle{Gradient Computation.} In Algorithm \ref{alg:bilevel_gradient_descent}, $\nabla_{\bm{w}}L_{train}(\bm{\beta} (\bm{\tau}), \bm{w})$ can be derived from Equation \ref{eqn:train_loss_param}, where $\bm{\beta}$ can be considered as constant. However, since $\bm{\beta}$ (Equation \ref{eqn:beta_definition}) and $\bm{w}'$ (Equation \ref{eqn:approx_optimal_model}) are variables of $\bm{\tau}$, to compute the gradients $\nabla_{\bm{\tau}}L_{val}(\bm{\beta}(\bm{\tau}), \bm{w}'(\bm{\tau}))$, we need to use the chain rule~\footnote{Chain rules for multivariable functions: Suppose that $x(t)$ and $y(t)$ are differentiable functions of $t$ and $z = f(x(t), y(t))$ is a differentiable function of $x$ and $y$. Then the chain rule states $\frac{dz}{dt} = \frac{\partial z}{\partial x} \frac{\partial x}{\partial t} + \frac{\partial z}{\partial y} \frac{\partial y}{\partial t}$ .}:
\begin{multline}
\nabla_{\bm{\tau}}L_{val}(\bm{\beta}(\bm{\tau}), \bm{w}'(\bm{\tau})) = \nabla_{\bm{\beta}}L_{val}(\bm{\beta}, \bm{w}') \cdot \nabla_{\bm{\tau}}\bm{\beta} \\- \eta_2 \nabla_{\bm{w'}}L_{val}(\bm{\beta}, \bm{w'}) \cdot  \nabla_{\bm{w}, \bm{\beta}} L_{train}(\bm{\beta}, \bm{w})   \cdot \nabla_{\bm{\tau}}\bm{\beta}
\label{eqn:grad_chain_rule}
\end{multline}
We can decompose Equation \ref{eqn:grad_chain_rule} into three parts: $D_1$, $D_2$, $D_3$, where 
\begin{align*}
&D_1 = \nabla_{\bm{\tau}}\bm{\beta} \\
&D_2 = \nabla_{\bm{\beta}} L_{val}(\bm{\beta}, \bm{w}') \\
&D_3 = \nabla_{\bm{w'}}L_{val}(\bm{\beta}, \bm{w'})\cdot \nabla_{\bm{w}, \bm{\beta}} L_{train}(\bm{\beta}, \bm{w})
\end{align*}
The gradient $D_1$ can be derived from Equation \ref{eqn:beta_definition}. The gradient $D_2$ is also easy to compute, where we consider $\bm{w}'$ as constant and compute the gradient of validation loss w.r.t. $\bm{\beta}$. This can be derived from Equation \ref{eqn:val_loss_param} or as we will show later, using backpropagation through the pipeline. However, computing $D_3$ is difficult since it involves a second-order derivative and matrix-vector product computation.  Following the previous work on neural network architecture search~\cite{liu2018darts}, we approximate it using numerical differentiation. Let $\epsilon$ be a small scalar and $\bm{w^{\pm}} = \bm{w} \pm \epsilon \nabla_{\bm{w'}}L_{val}( \bm{\beta}, \bm{w}')$, then $D_3$ can be estimated as:
\begin{align}
\begin{split}
    D_3 \approx \frac{\nabla_{\bm{\beta}}L_{train}( \bm{\beta}, \bm{w}^+)-\nabla_{\bm{\beta}}L_{train}(\bm{\beta}, \bm{w}^-)}{2\epsilon}
    \label{eqn:hessian_approx}
\end{split}
\end{align}
$\nabla_{\bm{\beta}}L_{train}(\bm{\beta}, \bm{w}^+)$, $\nabla_{\bm{\beta}}L_{train}( \bm{\beta}, \bm{w}^-)$ are the gradients of the training loss w.r.t $\bm{\beta}$, where we consider $\bm{w}^+$ and $\bm{w}^-$ as constant. They can be derived from Equation \ref{eqn:train_loss_param} or as we will show later using backpropagation through the pipeline.  Therefore, to update the underlying parameters $\bm{\tau}$, we need to compute three gradients w.r.t. $\bm{\beta}$ with different model parameters: $\nabla_{\bm{\beta}}L_{train}(\bm{\beta}, \bm{w}^+)$, $\nabla_{\bm{\beta}}L_{train}( \bm{\beta}, \bm{w}^-)$ and $\nabla_{\bm{\beta}} L_{val}(\bm{\beta}, \bm{w}')$. 


\stitle{Backpropagation through the pipeline.} The gradients of loss w.r.t. $\bm{\beta}$ can be computed using backward propagation through the pipeline. Let $\frac{\partial L}{\partial x_{s}}$ be the derivative of the loss w.r.t. the input of the ML model or the final output of the preprocessing pipeline. From Equation \ref{eqn:forward_prop}, using the chain rule, we have
\begin{align}
\begin{gathered}
    \frac{\partial L}{\partial {\beta_{ij}}} = \frac{\partial L}{\partial x_{i}} f_{ij}(x_{i-1}), \quad \frac{\partial L}{\partial {x_{i-1}}} = \frac{\partial L}{\partial x_{i}} \sum_{j=1}^m\beta_{ij}\frac{\partial f_{ij}}{\partial x_{i-1}}
\end{gathered}
\label{eqn:backpropagation}
\end{align}
 However, in this equation, the gradient term $\frac{\partial f_{ij}}{\partial x_{i-1}}$, namely, the gradient of the output of a TF  operator w.r.t. its input depends on the internals of a TF operator, which  may not be easy to compute. Also, we expect users to add their own customized TF operators, but users may not be able to derive the gradients. Therefore, we assume that all TF operators are black-box functions, where we only have access to the output and input without knowing their internal algorithms, and we approximate the gradients using numerical differentiation. Let $\epsilon$ be a small scalar. This gradient can be estimated as:
\begin{equation}
    \frac{\partial f_{ij}}{\partial x_{i-1}} \approx \frac{f_{ij}(x_{i-1} + \epsilon) - f_{ij} (x_{i-1} - \epsilon)}{2\epsilon}
    \label{eqn:numerical_diff}
\end{equation}
Then, we can use Equation \ref{eqn:backpropagation} to backpropagate the gradients of the loss w.r.t. the output of each TF  operator and each $\beta_{ij}$.

\stitle{Implementation with automatic differentiation.} Many  automatic differentiation engines (e.g. Pytorch~\cite{paszke2017automatic}, TensorFlow~\cite{abadi2016tensorflow}) can perform backpropagation automatically and efficiently to compute the gradients. However, these tools usually require that all the computations in the forward propagation are differentiable and implemented using their frameworks. Since we consider TF operators as black-box functions of which the internal algorithms are unknown, we cannot leverage these tools directly to backpropagate the preprocessing pipeline. To solve this issue, we modify the original forward propagation of the pipeline  (Equation \ref{eqn:forward_prop}) to be:
\begin{equation}
    \label{eq:auto_diff}
    x_i = \sum_{j=1}^m {\beta_{ij}\tilde{o}_{ij}}  + x_{i-1} \sum_{j=1}^m \tilde{\beta}_{ij} \tilde{d}_{ij}  - \tilde{x}_{i-1}  \sum_{j=1}^m \tilde{\beta}_{ij} \tilde{d}_{ij}
\end{equation}
where $\tilde{o}_{ij} = f_{ij}(x_{i-1}) $ is the output of the black-box TF  operator; $\tilde{d}_{ij} = \frac{\partial f_{ij}}{\partial x_{i-1}}$ is the numerical derivative computed using Equation~\ref{eqn:numerical_diff}; $\tilde{x}_{i-1}$ is a \textit{snapshot} of $x_{i-1}$, which is a constant number that has the same value as $x_{i-1}$ but does not require gradient (e.g., $\tilde{x}_{i-1} = x_{i-1}$.\texttt{detach()} in PyTorch or  $\tilde{x}_{i-1}= \texttt{stop\_gradient}(x_{i-1})$ in TensorFlow); $\tilde{\beta}_{ij}$ is a snapshot of $\beta_{ij}$. Note that all the variables with tilde are constant numbers that do not have gradients. 

In the forward pass, Equation \ref{eq:auto_diff} yields exactly the same outputs as Equation \ref{eqn:forward_prop}, but it only requires the output of the TF operators and the numerical derivatives computed using the input and output of TF operators. Therefore, the internal implementation of TF operators is not involved, which enables the automatic backpropagation to be performed. In the backward pass, the automatic differentiation engines will compute the gradients as:
\begin{align}
\begin{gathered}
    \frac{\partial L}{\partial {\beta_{ij}}} = \frac{\partial L}{\partial x_{i}} \tilde{o}_{ij},\quad \frac{\partial L}{\partial {x_{i-1}}} = \frac{\partial L}{\partial x_{i}} \sum_{j=1}^m\tilde\beta_{ij}\tilde d_{ij}
\end{gathered}
\label{eq:diff_L_x_prev}
\end{align}
This yields exactly the  same gradients as Equation \ref{eqn:backpropagation} with the gradients of black-box TF operators replaced by approximate numerical gradients. Therefore, this allows us to backpropagate the preprocessing pipeline correctly and automatically using any automatic differentiation engines.

\stitle{Algorithm.} Algorithm \ref{alg:auto_prep_fixed_prototype} shows the pseudocode of \ourmethod~ with a fixed prototype. We first initialize the underlying pipeline parameters and model parameters (Line 1). Most TF operators need to be fitted before performing transformation. For example, standardization needs to compute the mean and standard deviation of the input data. Therefore, at each iteration, we first fit all the TF operators using the transformed training data (Line 3). Note that each TF operator should be fitted on its input training data, which is the output data of its previous step in the pipeline (i.e., the operator $f_{ij}$ should be fitted on $x_{i-1}$ over all training data). Since $x_{i-1}$ depends on $\bm{\beta}$ parameters (Equation \ref{eq:auto_diff}), it will be changed every iteration as $\bm{\beta}$ parameters are updated. Therefore, we need to refit TF operators at the beginning of every iteration. Then we can perform forward propagation using Equation \ref{eq:auto_diff} (Line 4) and backward propagation using automatic differentiation (Line 5). We compute the gradients needed for updating underlying pipeline parameters $\bm{\tau}$ using Equation \ref{eqn:grad_chain_rule} and Equation \ref{eqn:hessian_approx} (Line 6). We update the underlying pipeline parameters and model parameters alternatively using gradient descent (Line 7 - 8). This will be repeated until convergence. 
 \begin{small}
\begin{algorithm}[t]
\caption{\ourmethod-Fix}
\label{alg:auto_prep_fixed_prototype}
\begin{algorithmic}[1]
\REQUIRE Space of TF types and operators $S$, pre-defined prototype $\mathcal{T}$, training set $D_{train}$, validation set $D_{val}$
\ENSURE Optimal pipeline parameters  $\bm{\beta}$ and model parameters $\bm{w}$
\STATE{Initialize $\bm{\tau}$ and $\bm{w}$}
\WHILE{\textit{not converged}}
\STATE{Fit TF operators on the transformed training data}
\STATE{Forward Propagation: Compute $L_{train}(\bm{\beta}, \bm{w}^+)$, $L_{train}(\bm{\beta}, \bm{w}^-)$,\\ $L_{val}(\bm{\beta}, \bm{w}')$}
\STATE{Backward Propagation: Compute $\nabla_{\bm{\beta}}L_{train}( \bm{\beta}, \bm{w}^+)$, \\$\nabla_{\bm{\beta}}L_{train}( \bm{\beta}, \bm{w}^-)$, $\nabla_{\bm{\beta}}L_{val}(\bm{\beta}, \bm{w}')$}
\STATE{Compute $\nabla_{\bm{\tau}}L_{val}(\bm{\beta}, \bm{w}')$}
\STATE{Update $\bm{\tau}$: $\bm{\tau} = \bm{\tau}- \eta_1  \nabla_{\bm{\tau}} L_{val}(\bm{\beta}, \bm{w}')$}
\STATE{Update $\bm{w}$: $\bm{w} = \bm{w} - \eta_2\nabla_{\bm{w}}L_{train}( \bm{\beta} (\bm{\tau}), \bm{w})$}
\ENDWHILE

\RETURN $\bm{\beta}(\bm{\tau})$, $\bm{w}$
\end{algorithmic}
\end{algorithm}
 \end{small}

\stitle{Complexity.} The number of $\bm{\beta}$ parameters needed is $s \times m \times c$, where $s$ is the number of TF types, $m$ is the number of TF operators for each TF type and $c$ is the number of features. The running time is dominated by the forward propagation and backward propagation. Without \ourmethod, the training process only needs to update the model parameters (Line 8), which requires one pass of forward propagation and one pass of backpropagation. With \ourmethod, updating $\bm{\beta}$ needs three more passes of forward propagation (Line 4) and three more passes of backpropagation (Line 5). To improve the efficiency of the algorithm, we fit TF operators (Line 3) and update $\bm{\beta}(\bm{\tau})$ (Line 4 - 7) with a mini-batch randomly sampled from the training/validation set, instead of using all training/validation examples, which is similar to using stochastic gradient descent in place of gradient descent. We apply this strategy in our experiments.
\section{Data Preprocessing with Flexible Prototype} 
 \label{sec:autoprep_flex_prototype}

 \begin{figure*}[t]
     \centering
     \includegraphics[width=1.9\columnwidth]{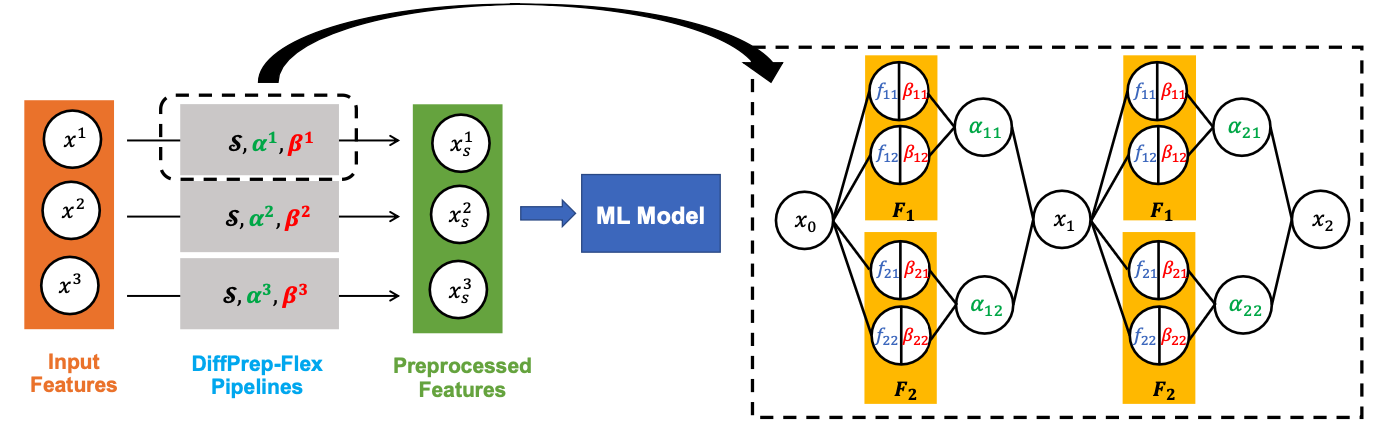}
     \vspace{-2mm}
     \caption{DiffPrep-Flex Architecture}
     \label{fig:autoprep_architect_flex}
 \end{figure*}

We now discuss the more general and flexible case, where the data preprocessing prototype is not pre-defined. In such a case, prior to operator selection using the method we introduced in Section \ref{sec:auto_prep_fixed_prototype}, we need to select a prototype from space for each feature. A prototype is an ordered sequence of TF types. Therefore, for prototype selection, we need to determine (1) the TF types that are included in the prototype and (2) the order of TF types in the sequence. 

To simplify this search problem, we introduce an \textit{identity transformation operator} ($I$) into each TF type, which is defined as $I(x) = x$. The identity TF operator is a function that maps data to itself, so selecting the identity TF operator for a TF type during operator selection is equivalent to dropping this TF type in the prototype. 
This allows us to include all TF types in the prototype and only focus on their order, rather than deciding which to include or exclude. A prototype becomes a permutation of all TF types with the option to exclude any by selecting the identity operator for it.

Now, the problem is: \textit{how to find the optimal prototype for each feature such that the optimal pipeline under the optimal prototype minimizes the validation loss?} A naive approach is to enumerate every possible prototype and use the method introduced in Section \ref{sec:auto_prep_fixed_prototype} to find the optimal pipeline for each prototype. However, this approach is not feasible because the search space of prototypes is large. Given a space that contains $s$ TF types, the number of possible permutations is $s!$. Since different features can use different prototypes, for a dataset with $c$ features, there are $s!^c$ possible combinations, which is exponential to the number of features. 

The hardness of the above search problem is again because the search space of prototypes is discrete and non-differentiable. To solve this problem efficiently, we use a method similar to that introduced in Section \ref{sec:auto_prep_fixed_prototype}: we first parameterize the search space using a set of binary prototype parameters (Section~\ref{sec:flex_param}) and then make the search space to be continuous and differentiable (Section~\ref{sec:flex_relax}). Finally, using the method introduced in Section \ref{sec:fixed_optimization}, we can solve the bi-level optimization using gradient descent, which enables our method to learn the prototype parameters, pipeline parameters and model parameters  simultaneously and efficiently in one training loop.

\subsection{Parameterization}
\label{sec:flex_param}
By introducing identity TF operators,  given a space of TF types $\mathcal{S} = \{F_1, F_2, ... F_s\}$, a prototype $\mathcal{T} = \{T_1, T_2, ... T_s\}$ would be a permutation of $\mathcal{S}$. Hence, we can represent a prototype  using a $s \times s$ \textit{permutation matrix} $\bm{\alpha} = \{\alpha_{ij}\}, \alpha_{ij} \in \{0, 1\}$ defined as follows. 
\begin{equation}
    \alpha_{ij} =
    \begin{cases}
        1 & \text{$T_i = F_j$}\\
        0 & \text{Otherwise}
    \end{cases}
    \label{eqn:def_alpha}
\end{equation}
In other words, $\alpha_{ij}$ is 1 if the $i$-th TF type in the prototype is $F_j$, otherwise it is 0. Note that the permutation matrix has exactly one element of 1 each row and each column and 0s elsewhere, i.e., the sum of each row and each column is exactly 1. Hence, we have $ \sum_i \alpha_{ij} = 1$ and $ \sum_j \alpha_{ij} = 1$.

\begin{example}
Assume that the space consists of three TF types $S = \{F_1, F_2, F_3\}$. Consider the prototype  $\mathcal{T} = \{F_2, F_3, F_1\}$. We have $\alpha_{12}=1, \alpha_{23}=1, \alpha_{31}=1$ and 0s elsewhere. Hence, This prototype can be represented using the following $\bm{\alpha}$ matrix.
\begin{equation*}
\bm{\alpha} = \begin{bmatrix}
0 & 1 & 0\\ 
0 & 0 & 1\\ 
1 & 0 & 0
\end{bmatrix}
\end{equation*}
\end{example}
\vspace{-1mm}

To generate a pipeline, in addition to the prototype, we also need to select a TF  operator for each TF type. Let $F_{i} = \{f_{i1}, f_{i2} ... f_{im}\}$, where $f_{ij}$ is the $j$-th TF  operator for the TF type $F_i$. Then, we can still use Equation \ref{eqn:def_beta} to define the $\bm{\beta}$ matrix, which represents the result of operator selection.  An $\bm{\alpha}$ matrix and a $\bm{\beta}$ matrix together uniquely defines a data preprocessing pipeline. The data after $i$ steps transformation can be computed as:
\begin{equation}
\label{eqn:forward_prop_flex}
x_i = \sum^s_{j=1}  \sum^m_{k=1} \alpha_{ij} \beta_{jk} f_{jk}(x_{i-1})
\end{equation}
Note that in Equation \ref{eqn:forward_prop_flex}, only the TF type $F_j$ selected for the $i$-th step has $\alpha_{ij}=1$ and only the TF operator selected for $F_j$ has $\beta_{jk} = 1$. Therefore, the summation is exactly the output of the selected TF  operator for the $i$-th TF type in the prototype. 
 
Similar to Equation \ref{eqn:train_loss_param} and \ref{eqn:val_loss_param}, we can also compute the training loss $L_{train}(\bm{\alpha}^1, ...,\bm{\alpha}^c, \bm{\beta}^1, ...,\bm{\beta}^c, \bm{w})$ and validation loss $L_{val}(\bm{\alpha}^1, ...,\bm{\alpha}^c, \\ \bm{\beta}^1, ...,\bm{\beta}^c, \bm{w})$ using Equation \ref{eqn:forward_prop_flex}, which will be parameterized by $\bm{\alpha}$, $\bm{\beta}$ matrices. We can also rewrite the DPPS problem statement using $\bm{\alpha}$, $\bm{\beta}$ matrices as:
\begin{align}
\label{eqn:problem_statement_param_outer_flex}
\min_{\bm{\alpha}^1,..., \bm{\alpha}^c, \bm{\beta}^1,..., \bm{\beta}^c} \quad & L_{val}(\bm {\alpha}^1,... \bm {\alpha}^c, \bm {\beta}^1,... \bm {\beta}^c, \bm{w}^*)\\
\label{eqn:problem_statement_param_inner_flex}
\text{s.t.} \quad & \bm{w}^*  = \text{argmin}_{\bm{w}}  L_{train}(\bm{\alpha}^1, ... \bm{\alpha}^c, \bm{\beta}^1,..., \bm{\beta}^c, \bm{w}) 
\end{align}

 \stitle{Architecture.} Figure \ref{fig:autoprep_architect_flex} shows the architecture of DiffPrep with a flexible prototype. 
 For each input feature, DiffPrep will learn a set of $\bm{\alpha}$ and $\bm{\beta}$ parameters that defines a data preprocessing pipeline. The $\bm{\alpha}$ parameters define the prototype and the $\bm{\beta}$ parameters define the results of operator selection. The final output of DiffPrep-Flex pipelines will be the preprocessed features parameterized by $\bm{\alpha}$, $\bm{\beta}$ parameters, which will be fed into ML models.

\subsection{Differentiable Relaxation}
\label{sec:flex_relax}
We can reuse the method described in Section \ref{sec:fixed_relax} to relax $\bm{\beta}$ parameters. For $\bm{\alpha}$ matrix, similar to relaxing $\bm{\beta}$ matrix, we can relax $\bm{\alpha}$ to be continuous variables $\alpha_{ij} \in [0, 1]$, but we need to retain the constraints between parameters in the original binary matrix ($\sum_i \alpha_{ij} = 1, \sum_j \alpha_{ij} = 1$).
In other words,
 $\bm{\alpha}$ matrix requires to be a non-negative squared matrix with both rows and columns summing up to 1, which is so-called a \textit{doubly stochastic matrix} (DSM). To enforce that $\bm{\alpha}$ is a DSM, we generate it using \textit{Sinkhorn normalization}. Sinkhorn~\cite{sinkhorn1964relationship, sinkhorn1967concerning} showed that any non-negative square matrix can be converted into a DSM by repeatedly and alternatively normalizing its rows and columns. Cruz et al.~\cite{santa2017deeppermnet} introduced \textit{Sinkhorn Layer} that converts CNN predictions to a DSM using Sinkhorn normalizations. Following Adams et al.~\cite{adams2011ranking}, we define Sinkhorn normalization over any squared matrix $X$ as:
\begin{align*}
    &C_{ij}(X) = \frac{X_{ij}}{\sum_i X_{ij}}, R_{ij}(X) = \frac{X_{ij}}{\sum_j X_{ij}}, S(X) = \lim\limits_{l\rightarrow \infty} S^l(X) \\
    &S^{l}(X) =
    \begin{cases}
        X_{ij} & l=1\\
        R(C(S^{l-1}(X)) & l>1 
    \end{cases}
\end{align*}
where $R$ and $C$ are the row  and column normalization. $S(X)$ is the result of Sinkhorn normalization, which is a DSM.
 
Inspired by these work, we define $\bm{\alpha}$ matrix as  $\bm{\alpha} = S(\bm{\theta})$,
where  $\bm{\theta} = \{\theta_{ij}\} \in \mathbb{R}^{s \times s}_{+}$ are non-negative underlying parameters. Note that this normalization is a differentiable process, which allows us to compute the gradients w.r.t the underlying parameters.
By enforcing these constraints, we can interpret $\alpha_{ij}$ (Equation \ref{eqn:def_alpha}) as the probability that $F_j$ is the $i$-th transformation type in the prototype, and $\beta_{ij}$ (Equation \ref{eqn:def_beta}) as the probability that $f_{ij}$ is selected for $F_i$. In this sense,  Equation \ref{eqn:forward_prop_flex} will return the expected value of the transformed data, which can be used to compute the expected training and validation loss.  

\subsection{Bi-level Optimization}
Now we can use gradient descent to solve the bi-level optimization (Equation~\ref{eqn:problem_statement_param_outer_flex} and \ref{eqn:problem_statement_param_inner_flex}) and find the optimal prototype and pipeline represented by $\bm{\alpha}$ and $\bm{\beta}$ parameters, similar to Algorithm \ref{alg:auto_prep_fixed_prototype}.

\stitle{Algorithm.} Algorithm~\ref{alg:auto_prep_flex_prototype} shows the pseudocode of DiffPrep with flexible prototypes. Compared to  Algorithm \ref{alg:auto_prep_fixed_prototype}, the input only requires a space of TF types and TF operators, and does not need a pre-defined prototype. At each iteration, we compute the gradients of the validation loss  with respect to both $\bm{\alpha}$ and $\bm{\beta}$ using the method described in Section~\ref{sec:fixed_optimization}  (Line 4 - 6). Then we update the prototype parameters, pipeline parameters and model parameters (Line 7 - 8) until convergence.

 \begin{small}
\begin{algorithm}[t]
\caption{\ourmethod-Flex}
\label{alg:auto_prep_flex_prototype}
\begin{algorithmic}[1]
\REQUIRE Space of TF types and operators $S$, training set $D_{train}$, validation set $D_{val}$
\ENSURE Optimal prototype parameters $\bm{\alpha}$, pipeline parameters $\bm{\beta}$, and model parameters $\bm{w}$
\STATE{Initialize $\bm{\theta}$, $\bm{\tau}$ and $\bm{w}$}
\WHILE{\textit{not converged}}
\STATE{Fit TF operators on the transformed training data}
\STATE{Forward Propagation: Compute $L_{train}(\bm{\alpha}, \bm{\beta}, \bm{w}^+)$, \\$L_{train}(\bm{\alpha},\bm{\beta},\bm{w}^-)$, $L_{val}(\bm{\alpha},\bm{\beta}, \bm{w}')$}
\STATE{Backward Propagation: Compute $\nabla_{\bm{\alpha}}L_{train}( \bm{\alpha}, \bm{\beta}, \bm{w}^+)$,\\ 
$\nabla_{\bm{\alpha}}L_{train}( \bm{\alpha}, \bm{\beta}, \bm{w}^-)$,  $\nabla_{ \bm{\alpha}}L_{val}(\bm{\alpha},\bm{\beta}, \bm{w}')$, $\nabla_{\bm{\beta}}L_{train}(\bm{\alpha}, \bm{\beta}, \bm{w}^+)$, \\
$\nabla_{\bm{\beta}}L_{train}(\bm{\alpha},  \bm{\beta}, \bm{w}^-)$, $\nabla_{\bm{\beta}}L_{val}(\bm{\alpha}, \bm{\beta}, \bm{w}')$}
\STATE{Compute $\nabla_{\bm{\tau}}L_{val}(\bm{\alpha}, \bm{\beta}, \bm{w}')$, $\nabla_{\bm{\theta}}L_{val}(\bm{\alpha}, \bm{\beta}, \bm{w}')$}
\STATE{Update $\bm{\tau}, \bm{\theta}$: $\bm{\tau} = \bm{\tau}- \eta_1  \nabla_{\bm{\tau}} L_{val}(\bm{\alpha}, \bm{\beta}, \bm{w}')$, \\$\hspace{41pt}\bm{\theta} = \bm{\theta}- \eta_1  \nabla_{\bm{\theta}} L_{val}(\bm{\alpha}, \bm{\beta}, \bm{w}')$}
\STATE{Update $\bm{w}$: $\bm{w} = \bm{w} - \eta_2\nabla_{\bm{w}}L_{train}( \bm{\alpha},\bm{\beta}, \bm{w})$}
\ENDWHILE

\RETURN $\bm{\alpha}(\bm{\theta}), \bm{\beta}(\bm{\tau})$, $\bm{w}$
\end{algorithmic}
\end{algorithm}
\end{small}

\section{Experiments}
\label{sec:experiment}
We conduct extensive experiments to evaluate the effectiveness and efficiency of \textsc{DiffPrep}. All our experiments were performed on a \rev{single} machine with a 2.20GHz Intel Xeon(R) Gold 5120 CPU. 
The source code of our experiments is available at 
\url{https://github.com/chu-data-lab/DiffPrep}.

\subsection{Experimental Setup}
\label{sec:experiment_setup}
\vspace{-2mm}
\stitle{Datasets.} We evaluate our approaches on \rev{18 diverse} real-world datasets from the OpenML data repository~\cite{OpenML2013}. \rev{These datasets are frequently used in the AutoML and data cleaning literature~\cite{li2021cleanml, berti2019learn2clean, neutatz2022data, krishnan2017boostclean, ledell2020h2o}.} Table \ref{tab:test_acc} shows the size and \rev{number of label classes} of each dataset, as well as the number of missing values and outliers detected using the Z-score method (i.e., values more than 3 standard deviations away from the mean of the column are considered outliers). We randomly split each dataset into training/validation/test sets by a ratio of 60\%/20\%/20\%.

\stitle{Model.} For all preprocessing methods considered in the experiments, we use Logistic Regression as the downstream ML model, which is a classical differentiable model commonly used in practice and frequently considered  in the  literature~\cite{thornton2013auto, feurer2020auto, olson2016tpot, krishnan2019alphaclean}. 

\stitle{Search Space.} We consider four TF types in our experiments, including missing value imputation, normalization, outlier removal and discretization. For each TF type, we select several TF operators that are widely used in practice and provided by scikit-learn~\cite{scikit-learn} as shown in Table~\ref{tab:preprocessing_space}. For operators that have parameters, we discretize them with different parameters. For example, for the Z-score method, we consider Z-score(2), Z-score(3) and Z-score(4) as 3 different TF operators. We also add an identity operator to each TF type which amounts to a skipping operator. Since most TF operators provided by scikit-learn require the input data to contain no missing values, we do not add the identity operator to the missing value imputation. We enforce the missing value imputation to be the first TF type in the prototype so that missing values are imputed in the first step. In addition, many operators only accept numerical input data. Therefore, after missing value imputation, we transform all categorical features into numerical features using one-hot encoding. 

\stitle{Methods Compared.} We compare the following data preprocessing approaches.
\begin{itemize}[leftmargin=*]
    \item \textit{\ourmethod-Fix \rev{(DP-Fix)}.} This is our approach with a pre-defined fixed transformation order. We use the order \textit{\{missing value imputation, normalization, outlier removal, discretization\}}. Our method learns the best $\bm{\beta}$ parameters that define a pipeline for each feature. Note that different features can have different $\bm{\beta}$ parameters and thus can be preprocessed differently. 
    
    \item \textit{\ourmethod-Flex \rev{(DP-Flex)}.} This is our approach with a flexible transformation order. It will automatically learn $\bm{\alpha}$ parameters that define the transformation order and $\bm{\beta}$ parameters that define a pipeline under the learned order for each feature. 

    \item \textit{Default \rev{(DEF)}.} This is the default data preprocessing pipeline that is commonly used in practice and adopted in AutoML frameworks such as H2O~\cite{ledell2020h2o}. It first imputes numerical missing values with the mean value of the column and categorical missing values with the most frequent value of the column. Then, it normalizes each feature using standardization. \rev{Note that this default pipeline does not include outlier removal and discretization, which is the default setup for many existing AutoML frameworks~\cite{ledell2020h2o, Azure}.} All features use the same pipeline. 

    \item \textit{RandomSearch (RS).} This approach searches for a pipeline by training ML models with randomly sampled pipelines $N$ times and selecting the one with the best validation accuracy. We use the same transformation order as \ourmethod-Fix and sample one TF operator for each TF type from the space (Table ~\ref{tab:preprocessing_space}) to generate a sampled pipeline. All features are preprocessed using the same pipeline. We set the number of trials $N = 20$. 

    \item \textit{Auto-Sklearn (AS)} \cite{feurer2020auto}. This is an open-source AutoML package with automatic data preprocessing. Its built-in data preprocessor performs missing value imputation, removing low-variance features and normalization for numerical features, and performs category shift, missing value imputation,  minority coalescence and one-hot encoding for categorical features. \rev{The built-in data preprocessor already contains operators like category coalescence, category shift, TF-IDF encoding that are not included in our space}. It uses Bayesian optimization to search the optimal pipeline that will be used to preprocess all the features. The time limit for search is set to 1 hour and default settings are used for other configurations. AS also provides a feature processor that performs feature engineering operators, such as feature extraction and feature embedding, which are out of the scope of data preprocessing. \rev{Since we focus on data preprocessing and feature engineering is not performed for all other methods, for a fair comparison, we turn off the feature processor and only use the data processor.} \rev{We will report the results of AS with its feature processor later in Section \ref{sec:beyond_data_prep}, where we will show that our methods can also be combined with a feature processor. }
        
    \item \textit{Learn2Clean (LC)} \cite{berti2019learn2clean}. This is a reinforcement-learning-based data cleaning and preparation method proposed in the literature. It uses Q-learning to select the optimal preprocessing pipeline that maximizes the ML model performance. The TF types, operators and order are all flexible, but it only generates one pipeline to preprocess all the features in the same way. We modify the search space of TF operators to be the same as ours (Table \ref{tab:preprocessing_space}).

    \item \rev{\textit{BoostClean (BC)}~\cite{krishnan2017boostclean}. This is an automatic data cleaning approach that uses boosting to select an ensemble of data cleaning operators (or pipelines) from a pre-defined candidate set to maximize validation accuracy. BoostClean combines the ML models trained on different transformed data induced by different pipelines, thus it needs to train multiple models on different candidate pipelines. BoostClean does not support feature-wise pipelines. To generate candidate pipelines, we use the same transformation order as DiffPrep-Fix and randomly sample $50$ pipelines from the space (Table \ref{tab:preprocessing_space}). BoostClean is designed for binary classification but can be adapted for multi-class classification by breaking it down into multiple binary tasks using the one-vs-all method. We adopt this method in our experiments. We set its ensemble size $B = 5$, the best setup it reported. 
    }

\end{itemize}

\stitle{Training Process.} We use an SGD optimizer to optimize the model parameters and use an Adam optimizer to optimize the pipeline parameters. The learning rate is tuned using the validation set. 
The batch size is 512 and the number of training epochs is 1000. We keep track of the validation loss during training and report the result at the epoch with the minimum validation loss. The training and evaluation are implemented using PyTorch~\cite{paszke2017automatic}, which utilizes parallelism by default. 

\stitle{Evaluation Metrics.} Our goal is to automatically and efficiently compose a data preprocessing pipeline to maximize downstream ML model performance. Therefore, we use the test accuracy of the model and the end-to-end running time as the evaluation metrics.  

\subsection{Performance Comparison} 
\label{sec:performance_comparison}

\begin{table*}[ht]
\caption{Comparison of model test accuracy on \rev{18} real-world datasets using different data preprocessing pipelines. \textsc{\ourmethod} (DP-Fix and DP-Flex combined) achieves the best test accuracy on \rev{15 out of 18} datasets.}
\resizebox{\textwidth}{!}{
\begin{tabular}{l|ccccc|ccccccc|ccc}
\toprule
 & \multicolumn{5}{c|}{\textbf{Data Characteristics}} & \multicolumn{7}{c|}{\textbf{Test Accuracy}} & \multicolumn{3}{c}{\textbf{Ablation Study}} \\
\textbf{Dataset} & \textbf{\#Ex.} & \textbf{\#Feat.} & \textbf{\#Classes} & \textbf{\#MVs} & \textbf{\#Out.} & \textbf{DEF} & \textbf{RS} & \textbf{AS} & \textbf{LC} & \textbf{BC} & \textbf{DP-Fix} & \textbf{DP-Flex} & \textbf{\begin{tabular}[c]{@{}c@{}}DP-Fix \\ (w/o feat-wise)\end{tabular}} & \textbf{\begin{tabular}[c]{@{}c@{}}DP-Fix \\ (worst order)\end{tabular}} & \textbf{\begin{tabular}[c]{@{}c@{}}DP-Fix\\  (train-opt)\end{tabular}} \\
\midrule
abalone & 4177 & 9 & 28 & 0 & 200 & 0.24 & 0.243 & 0.216 & 0.186 & 0.168 & 0.238 & \textbf{0.255} & 0.237 & 0.238 & 0.243 \\
ada\_prior & 4562 & 15 & 2 & 88 & 423 & 0.848 & 0.844 & 0.853 & 0.816 & 0.848 & \textbf{0.854} & 0.846 & 0.851 & 0.848 & 0.849 \\
avila & 20867 & 11 & 12 & 0 & 4458 & 0.553 & 0.598 & 0.615 & 0.597 & 0.585 & \textbf{0.638} & 0.63 & 0.613 & 0.637 & 0.637 \\
connect-4 & 67557 & 43 & 3 & 0 & 45873 & 0.659 & 0.671 & 0.667 & 0.658 & 0.69 & \textbf{0.732} & 0.701 & 0.688 & 0.719 & 0.676 \\
eeg & 14980 & 15 & 2 & 0 & 209 & 0.589 & 0.658 & 0.657 & 0.641 & 0.659 & \textbf{0.678} & 0.677 & 0.662 & 0.648 & 0.675 \\
google & 9367 & 9 & 2 & 1639 & 109 & 0.586 & 0.627 & \textbf{0.664} & 0.549 & 0.616 & 0.645 & 0.641 & 0.621 & 0.601 & 0.64 \\
house & 1460 & 81 & 2 & 6965 & 617 & 0.928 & 0.938 & \textbf{0.945} & 0.812 & 0.928 & 0.932 & \textbf{0.945} & 0.942 & 0.914 & 0.938 \\
jungle\_chess & 44819 & 7 & 3 & 0 & 0 & 0.668 & 0.669 & 0.678 & 0.676 & 0.667 & \textbf{0.682} & \textbf{0.682} & 0.683 & 0.682 & 0.681 \\
micro & 20000 & 21 & 5 & 0 & 8122 & 0.564 & 0.579 & 0.584 & 0.582 & 0.561 & 0.586 & \textbf{0.588} & 0.576 & 0.586 & 0.584 \\
mozilla4 & 15545 & 6 & 2 & 0 & 290 & 0.855 & 0.922 & \textbf{0.931} & 0.854 & 0.93 & 0.923 & 0.922 & 0.921 & 0.854 & 0.923 \\
obesity & 2111 & 17 & 7 & 0 & 25 & 0.775 & 0.841 & 0.737 & 0.723 & 0.652 & 0.893 & \textbf{0.896} & 0.879 & 0.863 & 0.9 \\
page-blocks & 5473 & 11 & 5 & 0 & 1011 & 0.942 & 0.959 & \textbf{0.969} & 0.92 & 0.951 & 0.957 & 0.967 & 0.952 & 0.945 & 0.964 \\
pbcseq & 1945 & 19 & 2 & 1445 & 99 & 0.71 & 0.73 & 0.712 & 0.704 & 0.72 & 0.725 & \textbf{0.743} & 0.722 & 0.71 & 0.722 \\
pol & 15000 & 49 & 2 & 0 & 8754 & 0.884 & 0.879 & 0.877 & 0.737 & 0.903 & 0.904 & \textbf{0.919} & 0.888 & 0.882 & 0.896 \\
run\_or\_walk & 88588 & 7 & 2 & 0 & 8548 & 0.719 & 0.829 & 0.851 & 0.728 & 0.835 & 0.907 & \textbf{0.917} & 0.862 & 0.888 & 0.898 \\
shuttle & 58000 & 10 & 7 & 0 & 5341 & 0.964 & 0.996 & 0.998 & 0.997 & 0.997 & \textbf{0.998} & 0.997 & 0.998 & 0.998 & 0.996 \\
uscensus & 32561 & 15 & 2 & 4262 & 2812 & 0.848 & 0.84 & 0.851 & 0.786 & 0.848 & \textbf{0.857} & 0.852 & 0.853 & 0.858 & 0.857 \\
wall-robot-nav & 5456 & 25 & 4 & 0 & 1871 & 0.697 & 0.872 & 0.869 & 0.69 & 0.9 & 0.898 & \textbf{0.914} & 0.905 & 0.882 & 0.894 \\
\bottomrule
\end{tabular}
}
\label{tab:test_acc}
\end{table*}

\stitle{Accuracy Comparison.} Table \ref{tab:test_acc} shows the test accuracy of the ML model on \rev{18} real-world datasets with different preprocessing methods. We have the following observations from Table \ref{tab:test_acc}:
\begin{enumerate}[leftmargin=*]
\item Different preprocessing pipelines can lead to significantly different model performances. For example, on wall-robot-nav, using different pipelines, the accuracy differs by more than 20\%. 

\item \textsc{\ourmethod} (DP-Fix and DP-Flex combined) achieves the best test accuracy on \rev{15 out of 18} datasets. Our methods surpass the best baseline method by more than 1\% on 9 datasets. In particular, on run\_or\_walk, obesity and connect-4, our methods outperform the best baseline method by 6.6\%, 5.5\% and 4.2\%, respectively. This shows the effectiveness of \textsc{\ourmethod} and the significant improvement of model performance by using a larger search space. \rev{The reasons for gains vary by datasets. For example, connect-4 may benefit from cleaning outliers properly as it has a great number of outliers. On this dataset, we found DP-Fix selects Z-score methods for some features, while selecting skip operators for some other features. This shows the benefits of DiffPrep by using feature-wise pipelines, where we can clean true outliers in some features without affecting other features that may not contain true outliers. We will explain the contribution of each component of DiffPrep in more detail in the ablation study (Section \ref{eval:ablation}). } 

\item  \ourmethod-Fix (DP-Fix) and \ourmethod-Flex (DP-Flex) respectively outperform the best baseline methods on \rev{11 and 13} datasets. Comparing \ourmethod-Flex with \ourmethod-Fix, \ourmethod-Flex outperforms \ourmethod-Fix on \rev{9 out of 18} datasets, but the two methods differ by less than 2\% on \rev{17} datasets. \ourmethod-Flex does not significantly improve upon \ourmethod-Fix despite the larger search space, because the fixed order we used for DiffPrep-Fix is already the optimal order for many datasets. Hence, changing the order in \ourmethod-Flex may not lead to further improvement. Also, as the gradient descent is not always guaranteed to find the global optimum, it is possible to find a less optimal solution, \rev{ especially for DiffPrep-Flex which has a harder optimization problem due to its flexibility}. \rev{However, DiffPrep-Fix with default order is insufficient as it may not work well for all datasets.} As we will show in the ablation study, if the order is not set properly,  \ourmethod-Fix may perform worse, while \ourmethod-Flex can avoid such cases. \rev{In addition, the default order will be invalid if users add custom TF types.} The benefit of \ourmethod-Flex is that it does not require and is not affected by the pre-defined transformation order.  

\item RandomSearch (RS) performs better than Default on \rev{15 out of 18} datasets. Especially on wall-robot-nav and run\_or\_walk, the accuracy is improved by 17.5\% and 11\%, respectively. This indicates that using the same default preprocessing pipeline for all datasets is generally not a good strategy, despite its wide adoption in practice.

\item BoostClean (BC) performs worse than DiffPrep (DP-Fix and DP-Flex combined) on 17 out of 18 datasets. Comparing BC with RandomSearch (RS), on binary classification tasks, BoostClean (BC) outperforms RS on 6 out of 9 datasets. This is because BoostClean uses an ensemble of models and statistical boosting to improve model performance. Also, BoostClean has a larger candidate set with 30 additional pipelines compared to RandomSearch, increasing its chance of finding a better pipeline. However, on datasets with more than 2 classes, BoostClean only outperforms RandomSearch on 3 out of 9 datasets. Especially on obesity and abalone, BoostClean is 18.9\% and 7.5\% worse than RandomSearch. This is because BoostClean uses the one-vs-all method to handle multi-class classification, which may cause class imbalance issue and thus lead to worse model performances when the number of classes is large.

\item Although Auto-Sklearn and RandomSearch use different optimization methods, they perform close to each other on most datasets, where the difference is less than 2\% on \rev{14 out of 18} datasets. This is because they use a similar search space.  In comparison, our methods consider a larger space with feature-wise pipelines and flexible order, which further improves the performance.

\item Learn2Clean (LC) on average performs worst and it outperforms Default only on 6 datasets. Although Learn2Clean has a larger search space than other baseline methods, the Q-learning method is not effective to find a good pipeline and thus it performs worse than other methods.

\end{enumerate}

\begin{figure}[h]
    \centering
     \begin{subfigure}[t]{\columnwidth}
         \centering
         \includegraphics[width=0.7\columnwidth]{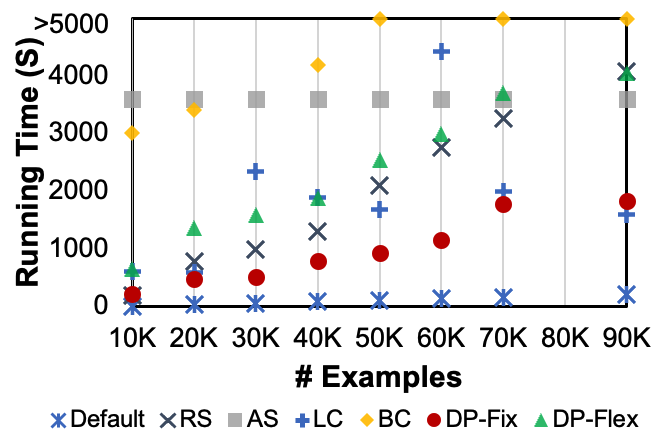}
         \vspace{-2mm}
         \caption{Logistic Regression}
         \label{fig:rev_time_log}
     \end{subfigure}
     \begin{subfigure}[t]{\columnwidth}
         \centering
         \includegraphics[width=0.7\columnwidth]
         {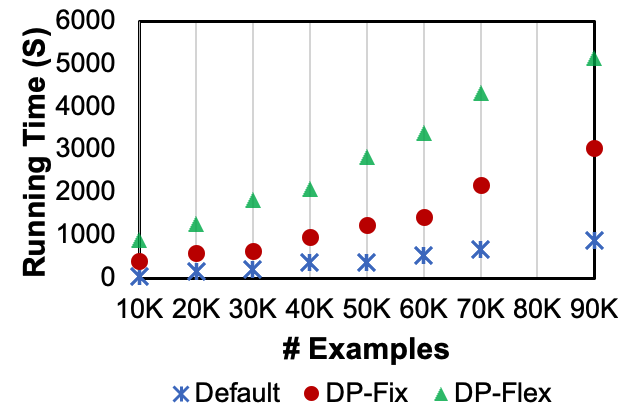}
         \vspace{-2mm}
         \caption{\rev{Two-layer Neural Network}}
         \label{fig:rev_time_two}
     \end{subfigure}
     \hfill
    \vspace{-2mm}
    \caption{Comparison of end-to-end running time of different preprocessing methods.}
    \label{fig:running_time}
\end{figure}

\stitle{Running Time Comparison}. To compare running time, we categorize all the datasets into different bins based on the number of rows of each dataset (e.g., 0-10K, 10-20K) and we report the average end-to-end running time of different methods over datasets in each bin as shown in Figure \ref{fig:running_time}(a).  As expected, Default takes the shortest time since it only trains the model once with the fixed pipeline. RandomSearch (RS) is about 20 times slower than Default, since it simply trains the model 20 times with 20 randomly sampled pipelines. Auto-Sklearn (AS) takes about 1 hour constantly as it keeps searching for the best result within the given 1 hour time limit. \rev{BoostClean (BC) is about 50 times slower than Default on binary class datasets as it needs to train a model on every candidate pipeline, but it is much slower on multi-class datasets as it uses the one-vs-all method and needs to rerun the whole algorithm for each subclass.} \rev{Additionally, while both RandomSearch and BoostClean can utilize parallelism to decrease running time by training multiple models simultaneously on multiple machines, this study only considers the single machine setup, which is more common for general users}. Overall, DiffPrep-Fix takes about half of the time of RandomSearch, which amounts to 10 times slower than Default and it is generally faster than Learn2Clean. DiffPrep-Flex takes similar time as RandomSearch and is generally faster than Auto-Sklearn. Also, we can see the running time of our methods grows linearly with the size of the dataset, showing their scalability.

Note that although DiffPrep only requires training the model once, the additional overhead comes from the forward/backward propagation through the pipelines. Recall that DiffPrep uses ``dynamic” pipelines that are changed at every iteration. Therefore, we need to reinvoke TF operators to transform data at every iteration. The ratio between our method and the Default method depends on the time it takes to propagate through the pipelines and through the model. Our experiments use logistic regression, in which propagation through the model is relatively fast. \rev{However, as the model sizes increase, the ratio between our methods and the Default will go down}. For example, 
Figure \ref{fig:running_time}(b) shows that with a larger ML model (two-layer neural network with 100 neurons), DiffPrep-Fix is 2-3 times slower and DiffPrep-Flex is 6-7 times slower than Default, both of which are much faster than other baseline methods. 

\subsection{Sensitivity Analysis}
\label{sec:experiment_sensitivity}

\stitle{Non-linear ML models.} One concern for our methods is that the approximate gradient computation (e.g.,  Equation~\ref{eqn:hessian_approx}) may not work well for non-linear ML models. To understand the impact of end ML models, we test our methods with a non-linear ML model: a two-layer neural network with ReLU activation and 100 neurons in the hidden layer. As shown in Table \ref{tab:two_layer_acc}, compared with other methods, \textsc{DiffPrep} (DP-Fix and DP-Flex combined) achieves the best test accuracy on 10 out of 18 datasets. Especially on pbcseq and avila, our method outperforms the best baseline method by 3.1\% and 3\%, respectively. This shows that our method is still effective  with non-linear models in practice. This corroborates with findings in DARTS~\cite{liu2018darts}, which uses similar approximate gradient computation for neural network architecture search and empirically works for non-linear models.

\begin{table}[ht]
\caption{\rev{Comparison of model test accuracy using a two-layer neural network.}}
\vspace{-3mm}
\scalebox{0.8}{
\begin{tabular}{l|ccccccc}
\toprule
\textbf{Dataset} & \textbf{DEF} & \textbf{RS} & \textbf{AS} & \textbf{LC} & \textbf{BC} & \textbf{DP-Fix} & \textbf{DP-Flex} \\
\midrule
abalone & 0.251 & 0.266 & 0.251 & 0.271 & 0.162 & 0.246 & \textbf{0.273} \\
ada\_prior & 0.829 & 0.845 & \textbf{0.855} & 0.815 & 0.829 & 0.834 & 0.84 \\
avila & 0.82 & 0.914 & 0.864 & 0.824 & 0.928 & 0.946 & \textbf{0.958} \\
connect-4 & 0.802 & 0.799 & 0.804 & 0.795 & 0.8 & \textbf{0.807} & 0.806 \\
eeg & 0.906 & 0.955 & 0.933 & 0.834 & \textbf{0.955} & 0.924 & 0.939 \\
google & 0.576 & 0.657 & 0.667 & 0.573 & 0.657 & \textbf{0.677} & 0.666 \\
house & 0.928 & 0.935 & 0.921 & 0.928 & \textbf{0.935} & 0.908 & 0.928 \\
jungle\_chess & 0.854 & 0.856 & 0.852 & 0.825 & 0.842 & 0.853 & \textbf{0.861} \\
micro & 0.626 & 0.626 & \textbf{0.63} & 0.617 & 0.615 & 0.628 & 0.626 \\
mozilla4 & 0.916 & 0.934 & 0.932 & 0.925 & 0.933 & \textbf{0.935} & 0.934 \\
obesity & 0.91 & \textbf{0.95} & 0.941 & 0.815 & 0.927 & 0.941 & 0.943 \\
page-blocks & 0.966 & 0.967 & 0.975 & \textbf{0.983} & 0.959 & 0.969 & 0.959 \\
pbcseq & 0.743 & 0.743 & 0.748 & 0.697 & 0.743 & \textbf{0.779} & \textbf{0.779} \\
pol & 0.989 & 0.967 & 0.988 & \textbf{0.99} & 0.988 & 0.981 & 0.984 \\
run\_or\_walk & 0.985 & 0.986 & 0.99 & 0.984 & 0.983 & 0.99 & \textbf{0.991} \\
shuttle & 0.999 & 0.998 & 0.999 & 0.999 & 0.999 & \textbf{1} & 0.999 \\
uscensus & 0.856 & 0.845 & 0.854 & 0.79 & \textbf{0.856} & 0.854 & 0.853\\
wall-robot-nav & 0.898 & 0.961 & 0.962 & 0.92 & 0.957 & \textbf{0.965} & 0.962 \\
\bottomrule
\end{tabular}
}
\label{tab:two_layer_acc}
\end{table}

\begin{figure}[h]
    \centering
    \includegraphics[width=\columnwidth]{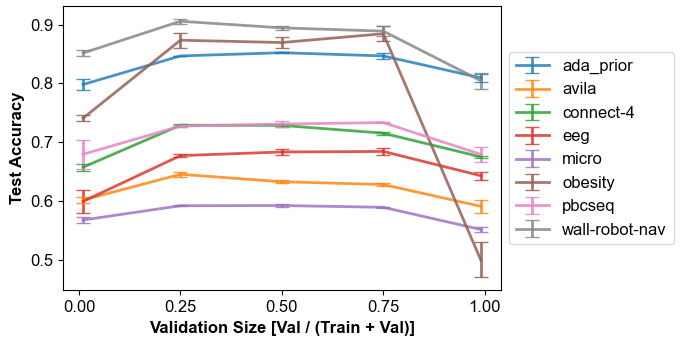}
    \vspace{-5mm}
    \caption{\rev{Using either a too small validation set or a too small training set can result in poor test accuracy due to overfitting. 60: 20 train/val split is generally good enough in practice.}}
    \label{fig:val}
\end{figure}

\stitle{Impact of validation size on overfitting.} \rev{Our methods aim to minimize validation loss, but this could lead to model overfitting on the validation set, especially when the validation set is small. To understand how the validation size affects the performance, we first hold out 20\% data as the test set and then randomly split the remaining data into validation/training sets with ratios ranging from 1\% to 99\%. We repeat it 5 times and run DiffPrep-Fix with each split. Figure \ref{fig:val} shows the mean and standard error of the test accuracy for some datasets. When the validation is extremely small (1\%), although the training set is large, the test accuracy is generally bad. This is because the pipeline parameters optimizing on the validation set make our method overfit the validation set. From 1\% to 25\%, the test accuracy improves significantly as overfitting is reduced. From 25\% to 50\%, the gain  becomes insignificant on most datasets because the validation set is large enough to prevent overfitting. However, as we further increase the validation size, the test accuracy drops on most datasets. Especially, when the validation set becomes extremely large (99\%), the accuracy is generally bad as the ML model overfits the small training set. Therefore, we can conclude that using either a too small validation set or a too small training set can result in overfitting issues and using 60:20 train/val split (i.e., 25\% ratio in the plot) is good enough in practice. The results on other datasets and on \ourmethod-Flex reveal similar findings and thus are  omitted here. }

\begin{table*}[ht]
\caption{Comparison of model accuracy and data cleaning quality on synthetic datasets with injected errors.}
\vspace{-3mm}
\resizebox{1.8\columnwidth}{!}{
\begin{tabular}{l|ccccccc|cccccc}
\toprule
\textbf{} & \multicolumn{7}{c|}{\textbf{Model Accuracy}} & \multicolumn{6}{c}{\textbf{Imputation Quality (RMSE)}} \\
\textbf{Dataset} & \textbf{DEF} & \textbf{RS} & \textbf{AS} & \textbf{LC} & \textbf{BC} & \textbf{DP-Fix} & \textbf{DP-Flex} & \textbf{DEF} & \textbf{RS} & \textbf{AS} & \textbf{LC} & \textbf{DP-Fix} & \textbf{DP-Flex} \\
\midrule
avila & 0.521 & 0.577 & 0.575 & 0.536 & 0.546 & \textbf{0.630} & 0.614 & 1.043 & 1.05 & 1.043 & 1.05 & \textbf{0.696} & 0.8 \\
eeg & 0.577 & 0.646 & 0.637 & 0.521 & 0.641 & 0.661 & \textbf{0.663} & \textbf{1787.601} & 1788.85 & \textbf{1787.601} & 1788.85 & 2662.249 & 2000.434 \\
wal-robot-nav & 0.66 & 0.863 & 0.818 & 0.646 & 0.862 & \textbf{0.892} & 0.888 & 1.25 & \textbf{0.965} & 1.25 & 1.312 & 1.273 & 1.427 \\
\bottomrule
\end{tabular}
}
\label{tab: synthetic}
\end{table*}

\begin{table*}[t]
\caption{\rev{Test accuracy combined with a feature preprocessor. }}
\vspace{-3mm}
\scalebox{0.8}{
\begin{tabular}{l|cc|cc}
\toprule
\multirow{2}{*}{\textbf{Dataset}} & \multicolumn{2}{c|}{\textbf{DP-Fix w/o Feat vs. w/ Feat}} & \multicolumn{2}{c}{\textbf{AS w/o Feat vs. w/ Feat}} \\

 & \textbf{\# Feats Change} & \textbf{Test Accuracy Change} & \textbf{\# Feats Change} & \textbf{Test Accuracy Change} \\
 \midrule
abalone & 9$\rightarrow$424 & 0.238$\rightarrow$0.26 & 9$\rightarrow$851 & 0.216$\rightarrow$0.237 \\
ada\_prior & 15$\rightarrow$408 & 0.854$\rightarrow$0.848 & 15$\rightarrow$120 & 0.853$\rightarrow$0.846 \\
avila & 11$\rightarrow$410 & 0.638$\rightarrow$0.994 & 11$\rightarrow$23760 & 0.615$\rightarrow$0.94 \\
connect-4 & 43$\rightarrow$445 & 0.732$\rightarrow$0.813 & 43$\rightarrow$10454 & 0.667$\rightarrow$0.808 \\
eeg & 15$\rightarrow$333 & 0.678$\rightarrow$0.93 & 15$\rightarrow$679 & 0.657$\rightarrow$0.91 \\
google & 9$\rightarrow$388 & 0.645$\rightarrow$0.673 & 9$\rightarrow$174 & 0.664$\rightarrow$0.674 \\
house & 81$\rightarrow$472 & 0.932$\rightarrow$0.935 & 81$\rightarrow$47 & 0.945$\rightarrow$0.928 \\
jungle\_chess & 7$\rightarrow$409 & 0.682$\rightarrow$0.809 & 7$\rightarrow$2106 & 0.678$\rightarrow$0.855 \\
micro & 21$\rightarrow$425 & 0.586$\rightarrow$0.605 & 21$\rightarrow$1771 & 0.584$\rightarrow$0.627 \\
mozilla4 & 6$\rightarrow$407 & 0.923$\rightarrow$0.947 & 6$\rightarrow$416 & 0.931$\rightarrow$0.94 \\
obesity & 17$\rightarrow$420 & 0.893$\rightarrow$0.943 & 17$\rightarrow$629 & 0.737$\rightarrow$0.886 \\
page-blocks & 11$\rightarrow$391 & 0.957$\rightarrow$0.967 & 11$\rightarrow$176 & 0.969$\rightarrow$0.969 \\
pbcseq & 19$\rightarrow$415 & 0.725$\rightarrow$0.807 & 19$\rightarrow$103 & 0.712$\rightarrow$0.692 \\
pol & 49$\rightarrow$442 & 0.904$\rightarrow$0.98 & 49$\rightarrow$623 & 0.877$\rightarrow$0.884 \\
run\_or\_walk & 7$\rightarrow$401 & 0.907$\rightarrow$0.989 & 7$\rightarrow$12942 & 0.851$\rightarrow$0.992 \\
shuttle & 10$\rightarrow$394 & 0.998$\rightarrow$0.999 & 10$\rightarrow$220 & 0.998$\rightarrow$0.925 \\
uscensus & 15$\rightarrow$416 & 0.857$\rightarrow$0.851 & 15$\rightarrow$279 & 0.851$\rightarrow$0.85 \\
wall-robot-nav & 25$\rightarrow$428 & 0.898$\rightarrow$0.994 & 25$\rightarrow$1956 & 0.869$\rightarrow$0.759 \\
\bottomrule
\end{tabular}
}
\label{tab:as}
\end{table*}

\subsection{Ablation Study}
\label{eval:ablation}

\stitle{Contribution of feature-wise pipelines.} To study the benefits of using feature-wise pipelines, we turn off feature-wise pipelines on DiffPrep-Fix by letting all features share the same $\bm{\beta}$ parameters. This way, our method can still learn the optimal $\bm{\beta}$ parameters but all features will be processed in the same way.  The result is shown in Table \ref{tab:test_acc} as DP-Fix (w/o feat-wise). Without feature-wise pipelines, the test accuracy drops on \rev{14 out of 18} datasets. Especially on run\_or\_walk and connect-4, the accuracy is reduced by 4.5\% and 4.4\%, respectively. This shows that using feature-wise pipelines can indeed improve the performance as it significantly enlarges the search space. Note that compared with the baseline methods in Table~\ref{tab:test_acc}, even without feature-wise pipelines, our method still achieves the highest accuracy on \rev{7 out of 18} datasets. This is because our method uses continuous parameters, which amounts to using a combination of different TF operators for each TF type and thus results in a larger space than choosing a single TF operator for each TF type.

\stitle{Benefits of flexible order.}  Compared with DiffPrep-Fix that uses a pre-defined transformation order, DiffPrep-Flex can search for the optimal order automatically with no need of user input. This is essential for less experienced users as they may not know how to set the order properly. To understand the benefit of using flexible order, we run DiffPrep-Fix with different transformation orders and we show the performance with the worst order for each dataset in Table \ref{tab:test_acc} as DP-Fix (worst order). With a bad order, the performance of DiffPrep-Fix drops significantly: on mozilla4, for example, the performance is reduced by 6.9\%. Compared to DiffPrep-Fix with the worst order, DiffPrep-Flex (DP-Flex) performs better on \rev{13 out of 18} datasets. This suggests that for inexperienced users, DiffPrep-Flex is a better method as it can avoid performance degradation due to bad pre-defined order.

\stitle{Alternative Optimization Objective.} To understand the benefits of bi-level optimization, we experiment with an alternative, one-level optimization objective, where both pipeline parameters ($\bm{\alpha}$, $\bm{\beta}$) and model parameters ($\bm{w}$) are learned to minimize the training loss. This optimization problem can be solved by computing the gradient of the training loss w.r.t each parameter and updating $\bm{\alpha}$, $\bm{\beta}$, $\bm{w}$ simultaneously using gradient descent. We run DiffPrep-Fix using this alternative objective and the results are shown in Table \ref{tab:test_acc} as DP-Fix (train-opt). Compared with this alternative method, our original DiffPrep-Fix (DP-Fix) performs better on 12 out of 18 datasets and has the same accuracy on 2 datasets.  We hypothesize that this is because optimizing pipelines and models jointly on training data makes it easier to overfit the training data and thus reduces the test accuracy. In addition, although our method achieves better accuracy in most cases, the difference between the two methods is less than 1\% on 17 out of 18 datasets. This is because even though the alternative objective only involves the training set, we tune hyperparameters (e.g., learning rates) using the validation set and adopt early stopping to mitigate the overfitting. Similar findings are also observed in DARTS~\cite{liu2018darts}, which uses a validation set to select network architecture. 

\subsection{Case study with Synthetic Datasets}
\label{sec:experiment_synthetic}

Traditional data cleaning or preprocessing methods focus on improving the data quality without considering downstream ML models. However, in the context of data cleaning for ML, prior works~\cite{li2021cleanml, neutatz2021cleaning} have shown that this may not improve and can even downgrade the ML model performance. This motivates us to take ML models into consideration and use the model accuracy as guidance to select preprocessing pipelines. To justify this motivation, we randomly inject 10\% missing values into three datasets. Since the ground truth of the synthetic datasets is known, we can evaluate the quality of missing value imputation using the root mean squared error (RMSE) between imputed values and ground truth values.
Table \ref{tab: synthetic} shows the model accuracy and the quality of missing value imputation using different preprocessing methods. Note that we omit data quality for BoostClean (BC) as it uses an ensemble of ML models trained on different imputed data.  The results show that DiffPrep consistently achieves the best model accuracy on all synthetic datasets.  However, the imputation methods selected by our methods are not necessarily the method with the best data quality. For example, on wal-robot-nav, the imputation quality of DiffPrep-Flex is the worst, however, it achieves better accuracy than other baseline methods. This indicates that selecting operators by data quality purely may not lead to the best model performance. Instead, we need to jointly consider other operators in the pipeline and the ML model to achieve the best result.

\subsection{Beyond Data Preprocessing}
\label{sec:beyond_data_prep}

\rev{While DiffPrep focuses on optimizing data preprocessing-related parameters, it can be easily integrated with AutoML methods that target other components in the ML pipeline, such as feature selection, feature embedding and hyperparameter tuning. In previous experiments, we have combined DiffPrep with learning rate tuning. In this example, we demonstrate how DiffPrep can work with feature extraction. Specifically, we train a random forest classifier on the raw input data as a features extractor, where we can use the tree node embeddings and predicted probability as the extracted features. We append the extracted features to the original data and feed the data into DiffPrep for data preprocessing and model training. Table~\ref{tab:as} shows the changes in the number of features and test accuracy of DP-Fix with and without the feature extractor. The accuracy significantly improved on 16 out of 18 datasets due to feature extraction. For reference, we report the performance change of Auto-Sklearn (AS) with its feature processor turned on, which performs feature extraction and selection on top of data preprocessing. The accuracy is significantly improved on 11 out of 18 datasets. Note that the performance on some datasets become worse, as enabling feature processor can make the search harder due to the increase in training time and search space.
}

\section{Related Work}
\label{sec:related_work}

\stitle{Automated Machine Learning.} The traditional workflow of developing ML models requires significant domain knowledge and human effort, especially for data preprocessing and model training, which can be time-consuming and expensive. Automated machine learning, also known as AutoML, aims to reduce the need of human involvement by automating the whole process of ML model development. Our work provides an effective and efficient solution to automate the process of data preprocessing.

Many existing AutoML tools enable automation in the process of data preprocessing and model training. Azure~\cite{Azure} is an AutoML system developed by Microsoft. It uses matrix factorization and Bayesian optimization to automatically select models and tune hyperparameters. However, for data preprocessing, it only considers tuning the normalization methods. For other transformations, it simply uses pre-defined default methods, such as mean imputation for missing value imputation. The order of transformations is fixed and different features are preprocessed in the same way. Auto-Sklearn~\cite{feurer2020auto} is an open-source AutoML package. It also relies on Bayesian optimization to automate data preprocessing and model training. Specifically, it adopts random-forest-based sequential model-based optimization to explore the search space. For data preprocessing, Auto-sklearn considers tuning a larger set of transformations than Azure. However, it also uses a fixed order of transformation and preprocesses all features in the same way. In comparison, our methods allow flexible order of transformations and different features can use different pipelines. Instead of Bayesian optimization, we use bi-level optimization with gradient descent, which is faster and only needs to train the model once.

\stitle{\rev{Data Cleaning for ML.}} \rev{Our work builds upon a line of data cleaning methods that incorporate signals from the downstream ML models into the design of cleaning objectives~\cite{neutatz2021cleaning}.} \rev{DiffML~\cite{hilprecht2022diffml} is our closest work in this area since it adopts a similar idea of making ML pipelines differentiable so that preprocessing steps can be jointly learned with the ML model using backpropogation. However, DiffML considers each preprocessing step separately while we consider a pipeline of preprocessing steps and formulate the ordering using Sinkhorn. In addition, DiffML minimizes training loss with one-level optimization, while we minimize validation loss with bi-level optimization, which reduces the risk of overfitting. } BoostClean~\cite{krishnan2017boostclean} automatically selects cleaning algorithms from the search space via statistical boosting to maximize the ML model's validation accuracy. \rev{It supports conditional cleaning operations defined by a combination of custom detection and repair functions. However, only 192 to 1976 operators were used in the evaluation, which is much smaller than the number of pipelines evaluated in our search space. BoostClean does not consider feature-wise pipelines.}  \rev{AlphaClean~\cite{krishnan2019alphaclean} is a similar automatic data cleaning pipeline generation system that finds cleaning pipelines from data cleaning libraries to maximize user-defined quality measures. AlphaClean uses tree search algorithms and learns pruning heuristics to reduce the search space. } Learn2Clean~\cite{berti2019learn2clean} uses reinforcement learning to select a sequence of data preprocessing operators such that the ML model performance is maximized. However, reinforcement learning needs to train and evaluate models many times to obtain rewards, which could lead to scalability challenges. ActiveClean~\cite{krishnan2016activeclean} focuses on gradient-based models and prioritizes cleaning examples with higher gradients that are likely to have large impacts on the model. CPClean~\cite{karlavs2020nearest} quantifies the impact of data cleaning on ML models using the uncertainty of predictions and prioritizes cleaning examples that would lead to the maximum reduction on the uncertainty of predictions. Both ActiveClean and CPClean solve the problem of which examples should be cleaned, while our method is design to select appropriate preprocessing/cleaning operators. \rev{Data Cleaning and AutoML~\cite{neutatz2022data} investigated the impact of data cleaning for AutoML systems and found that data errors did not affect the AutoML performance significantly as AutoML can select robust models and adjust ML pipelines to handle data errors properly. This verifies our motivation to take downstream models into consideration when handling data errors rather than isolating data cleaning as a separate process. }

\rev{\stitle{Network Architecture Search.} Network architecture search (NAS) aims to automate the design of neural network architectures. Although our work is not directly related to NAS, many of our ideas are inspired by DARTS~\cite{liu2018darts}, which is a differentiable NAS method. To search network architecture efficiently, DARTS first relaxes the categorical choice of network operators using continuous architecture parameters, which is similar to the pipeline parameters we used for choices of TF operators. 
It then formulates the architecture search as a bi-level optimization problem, which can be solved by updating the architecture and model parameters alternatively using gradient descent. Although DARTS and DiffPrep are similar in their methodology, there are two major differences. First, most architecture operators (e.g., convolution, ReLU) are differentiable and their gradients can be easily computed using automatic differentiation engines (e.g., PyTorch). However, the TF operators for data preprocessing can be complex and the gradients may not be easily computed. Second, the same type of TF operators rarely appears more than once in a pipeline and the order of TF operators (types) are usually flexible. In comparison, architecture operators usually repeat multiple times in an architecture but their order is usually fixed (e.g., DARTS uses ReLU-Conv-BN).
}

\section{Conclusion}
\label{sec:conclusion}

In this paper, we propose \textsc{\ourmethod}, a method that can automatically and efficiently search the optimal data preprocessing pipeline for a given tabular dataset and an ML model such that the performance of the ML model is maximized. We formalize the problem of automatic data preprocessing as a bi-level optimization problem. We then relax the discrete search space using continuous parameters, which enables us to search optimal preprocessing pipelines efficiently using gradient descent. The experiments show that our method achieves the best test accuracy on \rev{15 out of 18} real-world datasets and improves the model's test
accuracy by up to 6.6 percentage point.  We note that our proposed differentiable search strategy can only search for optimal data preprocessing pipelines when the end model is differentiable. We leave it to future work on how to search for the optimal data processing pipeline for non-differentiable end models, such as random forests. 

\begin{acks}
We thank the many members of the Georgia Tech Database Group for their valuable feedback on this work. This research was supported in part by Bosch Research North America.
\end{acks}

\bibliographystyle{ACM-Reference-Format}
\bibliography{sample-base}

\appendix

\end{document}